\documentclass{aa}

\usepackage{float} 
\usepackage{capt-of} 
\usepackage[toc,page,titletoc]{appendix} 

\usepackage{textcomp}

\usepackage{amsmath, amssymb, amsfonts, amscd}
\usepackage{mathtools} 
\usepackage{array} 
\usepackage{cases} 
\usepackage{empheq} 
\usepackage{slashbox} 
\usepackage{multirow} 
\usepackage{txfonts} 

\usepackage{tabularx,array}
\newcolumntype{G}{!{\vrule width 1pt}} 

\usepackage{graphicx} 
\usepackage{wrapfig} 

\usepackage{color} 
\usepackage{ulem} 
\usepackage[colorlinks=true,urlcolor=blue]{hyperref} 

\usepackage[toc,page,titletoc]{appendix}

\DeclareSymbolFont{boldsymbols}{OMS}{cmsy}{b}{n}
\DeclareSymbolFontAlphabet{\mathbfcal}{boldsymbols}

\newcommand{\mcal}{\mathcal}

\newcommand{\mbf}{\mathbf}
\newcommand{\mrm}{\mathrm}

\renewcommand{\Re}{\mathcal R \!e \ }
\renewcommand{\Im}{\mathcal I\!m \ }

\newcommand{\se}{\geqslant}

\newcommand{\nab}{{\pmb\nabla}}

\newcommand{\tenseur}[1]{ {\bar{ \bar{\pmb #1} }} }
\newcommand{\tenseurcomplex}[1]{ {\stackrel{\approx}{\pmb #1} }}

\newcommand{\intervalle}[3]
 { #1 \in \textrm{\textlbrackdbl}#2,#3\textrm{\textrbrackdbl} }
\newcommand{\interval}[2]
 { \textrm{\textlbrackdbl}#1,#2\textrm{\textrbrackdbl} }

\begin{document}

\title{Anelastic tidal dissipation in multi-layer planets}

\author{F. Remus\inst{1, 2, 3}, S. Mathis\inst{3, 4}, J.-P. Zahn\inst{1}, V. Lainey\inst{2}}
\offprints{F. Remus}

\institute{
 LUTH, Observatoire de Paris -- CNRS -- Universit\'e Paris Diderot, 5 place Jules Janssen, F-92195 Meudon Cedex, France
 \and
 IMCCE,  Observatoire de Paris -- UMR 8028 du CNRS -- Universit\'e Pierre et Marie Curie, 77 avenue Denfert-Rochereau, F-75014 Paris, France
 \and
 Laboratoire AIM Paris-Saclay, CEA/DSM -- CNRS -- Universit\'e Paris Diderot, IRFU/SAp Centre de Saclay, F-91191 Gif-sur-Yvette, France
 \and
 LESIA, Observatoire de Paris -- CNRS -- Universit\'e Paris Diderot -- Universit\'e Pierre et Marie Curie, 5 place Jules Janssen, F-92195 Meudon, France
 \\{}\\{}
 \email{francoise.remus@obspm.fr, stephane.mathis@cea.fr, jean-paul.zahn@obspm.fr, lainey@imcce.fr} 
}

\date{Received December 6, 2011; accepted February 24, 2012}

\abstract
{Earth-like planets have viscoelastic mantles, whereas giant planets may have viscoelastic cores.
The tidal dissipation of such solid regions, gravitationally perturbed by a companion body, highly depends on their rheology and on the tidal frequency. 
Therefore, modelling tidal interactions presents a high interest to provide constraints on planets' properties and to understand their history and their evolution, in our Solar Sytem or in exoplanetary systems.
}
{
The purpose of this paper is to examine the equilibrium tide in the anelastic parts of a planet whatever the rheology, taking into account the presence of a fluid envelope of constant density. 
We show how to obtain the different Love numbers that describe its tidal deformation. 
Thus, we discuss how the tidal dissipation in solid parts depends on the planet's internal structure and rheology. 
Finally, we show how the results may be implemented to describe the dynamical evolution of planetary systems.
}
{ 
We expand in Fourier series the tidal potential, exerted by a point mass companion, and express the dynamical equations in the orbital reference frame. The results are cast in the form of a complex disturbing function, which may be implemented directly in the equations governing the dynamical evolution of the system.
}
{
The first manifestation of the tide is to distort the shape of the planet adiabatically along the line of centers. 
Then, the response potential of the body to the tidal potential defines the complex Love numbers whose real part corresponds to the purely adiabatic elastic deformation, while its imaginary part accounts for dissipation. 
The tidal kinetic energy is dissipated into heat through anelastic friction, which is modeled here by the imaginary part of the complex shear modulus. 
This dissipation is responsible for the imaginary part of the disturbing function, which is implemented in the dynamical evolution equations, from which we derive the characteristic evolution times.
}
{
The rate at which the system evolves depends on the physical properties of tidal dissipation, and specifically on how the shear modulus varies with tidal frequency, on the radius and also the rheological properties of the solid core.
The quantification of the tidal dissipation in solid cores of giant planets reveals a possible high dissipation which may compete with dissipation in fluid layers. 
}

\keywords{ 
 Planet-Star interactions
 \,--\,
 Planets and satellites: general
 \,--\, 
 Planets and satellites: physical evolution
 \,--\, 
 Planets and satellites: dynamical evolution and stability
}

\titlerunning{Anelastic tidal dissipation in multi-layer planets}
\authorrunning{F. Remus, S. Mathis, J.-P. Zahn \& V. Lainey}

\maketitle

\section{Introduction and general context}
\label{section:intro}

Since 1995 a large number of extrasolar planets have been discovered, which display a wide diversity of physical parameters (Santos \&  et al., 2007).
Quite naturally the question arose of their habitability, i.e. whether they could allow the development of life. 
Determining factors are the presence of liquid water and of a protective magnetic field, properties which are closely linked to the values of the rotational and orbital elements of the planetary systems. 
And these elements strongly depend on the action of tides.
Once a planetary system is formed in a turbulent accretion disk, its fate is determined by the initial conditions and the mass ratio between planet and hosting star.
Through tidal interaction between components, the system evolves either to a stable state of minimum energy (where all spins are aligned, the orbits are circular and the rotation of each body is synchronized with the orbital motion) or the companion tends to spiral into the parent body.
Indeed, by converting kinetic energy into heat through internal friction, tidal interactions modify the orbital and rotational properties of the components of the considered system, and thus their structure through internal heating.
This mechanism depends sensitively on the internal structure and dynamics of the perturbed body.

Recent studies have been carried out on tidal effects in fluid bodies like stars and envelopes of giant planets (Ogilvie \&  Lin 2004-2007; Ogilvie 2009; Remus, Mathis \&  Zahn 2012).
However the planetary solid regions, such as mantles of Earth-like planets or rocky cores of giant planets, if present (e.g. Guillot 1999, Gaulme et al. 2011), may contribute likewise to tidal dissipation.
The first study of a tidally deformed elastic body was done by Lord Kelvin (1863) who applied it to an incompressible homogeneous Earth.
Further developments were made by Love (1911), who introduced a set of dimensionless numbers, the so-called Love numbers, to quantify the tidal perturbation. 
More recently Greff-Lefftz (2005) generalized these results in the case of a spheroidal rotating Earth. 
In the meanwhile Dermott (1979) considered a two-layer model and he studied the impact of a tidally deformed static fluid shell on the deformation of an elastic solid core.

If the body is not perfectly elastic, i.e. if its internal structure is anelastic, the tidal deformation presents a lag with respect to the tension exerted by the tidal force, and  causes the dissipation of kinetic energy.
Several recent studies addressed this problem of tidal dissipation using linear viscoelastic models.
Peale \&  Cassen (1978) evaluated the tidal dissipation in the Moon considering various models of internal structure. 
Tobie et al. (2005) applied the Maxwell rheological model to evaluate the dissipation in Titan and Europa; Ross \&  Schubert (1986) compared three different linear models of viscoelasticity (Kelvin-Voigt, Maxwell, Standard Anelastic Solid), to which Henning et al. (2009) added the Burgers body.
All these studies suscitate our interest in the tidal dissipation resulting from the anelastic deformation of the solid parts of a planet when perturbed by a companion.

We shall study here the tidal dissipation in a planet which possesses an anelastic core, made of a mix of ice and rock, surrounded by a fluid envelope, such as an ocean. 
The planet is part of a binary system where what we call the companion (or perturber) may be either the hosting star or a satellite of the planet. 
Due to the tide exerted by the companion, the core of the two-layer planet is deformed elastically, but because of the anelasticity of the material composing the core, this deformation is accompanied by a viscous dissipation that we evaluate whatever the rheology. 
As an illustration, the results will be given for a Maxwell body. 
We then compare the value of the tidal dissipation in presence of a fluid envelope with that achieved by the fully solid planet, and we examine the dependence of the results on the relative sizes of the core and the planet, the relative densities, and the viscoelastic parameters. 
In the last section, we establish the equations governing the dynamical evolution of the system, from which we deduce the caracteristic times of circularization, synchronization and spin alignments.

\section{Elastic deformations of a solid body under tidal perturbation}

\subsection{The system}

\begin{figure}[!htb]
 \centering
 \includegraphics[width=0.5\textwidth] {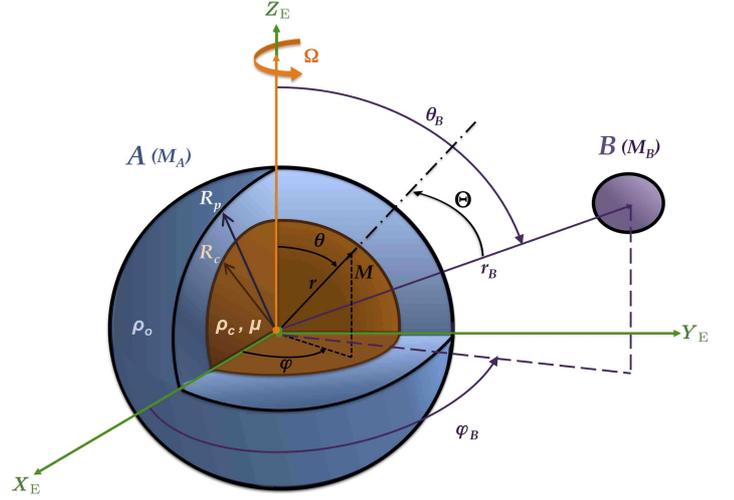}
 \caption{
 Spherical coordinates system attached to the equatorial reference frame ${\mcal{R}_E: \{ A, \bf X_E, \bf Y_E, \bf Z_E \}}$ associated to body $A$. A point $M$ is located by ${\mbf r \equiv \left( r,\theta,\varphi \right)}$; the point mass body $B$ by ${\mbf r_B \equiv \left( r_B,\theta_B,\varphi_B \right)}$.
 \label{fig:Fig_Bicouche}
 }
\end{figure}

We consider a planet $A$ of mass $M_A$, consisting of a rocky (or icy) core and a fluid envelope, rotating at the angular velocity $\Omega$ and tidally perturbed by a second body of mass $M_B$, assumed to be ponctual, moving around $A$ on a Keplerian orbit, of semi-major axis $a$ and eccentricity $e$, at the mean motion $n$.
We locate any point $M$ in space by its usual spherical coordinates ${\left( r, \theta, \varphi \right)}$ in a spin equatorial reference frame ${\mathcal{R}_E: \{ A, \bf X_E, \bf Y_E, \bf Z_E \}}$ centered on body $A$ and whose axis ${\left(A,\bf Z_E\right)}$ has the direction of the rotation axis of $A$.
The corresponding configuration is illustrated on Figure \ref{fig:Fig_Bicouche}.

\vspace{0.5cm}

In this section, we first assume that planet $A$ has no fluid layer and its internal structure is supposed to be perfectly elastic.
We will then denote by $\rho$ its density and $R$ its mean radius.

\subsection{The tidal potential}
\label{section:potU}

The planet is submitted to a tidal force, exerted by the perturber $B$, which derives from a perturbing time dependent potential $U\left(\mbf r,t\right)$.
Following Zahn (1966a-b) and generalizing it by using Kaula (1962), Lambeck (1980), Yoder (1995-1997) and Mathis \& Le Poncin (2009) (hereafter MLP09) in the present case of a close binary system where spins are not aligned, the components are not synchronized with the orbital motion and where the orbit is not circular, we expand the tidal potential $U$ in spherical harmonics ${Y_l^m(\theta,\varphi)}$ in $\mathcal{R}_E$.

Before we proceed, we need to define the Euler angles that link the spin equatorial frame ${\mathcal{R}_E: \{ A, \bf X_E, \bf Y_E, \bf Z_E \}}$ of the central body $A$, on one hand, and the orbital frame ${\mathcal{R}_O: \{ A, \bf X_O, \bf Y_O, \bf Z_O \}}$, on the other hand, to the quasi-inertial frame ${\mathcal{R}_R: \{ A, \bf X_R, \bf Y_R, \bf Z_R \}}$ whose axis $\bf Z_R$ has the direction of the total angular momentum of the whole system.

We need the three following Euler angles to locate the orbital reference frame $\mathcal{R}_O$ with respect to $\mathcal{R}_R$:
\vspace{-6pt}
\begin{itemize}
	\item $I$, the inclination of the orbital plane of $B$;
	\item $\omega^*$, the argument of the orbit pericenter;
	\item $\Omega^*$, the longitude of the orbit ascending node.
\end{itemize}
\vspace{-6pt}

\begin{figure}[!htb]
 \centering
 \includegraphics[width=0.5\textwidth] {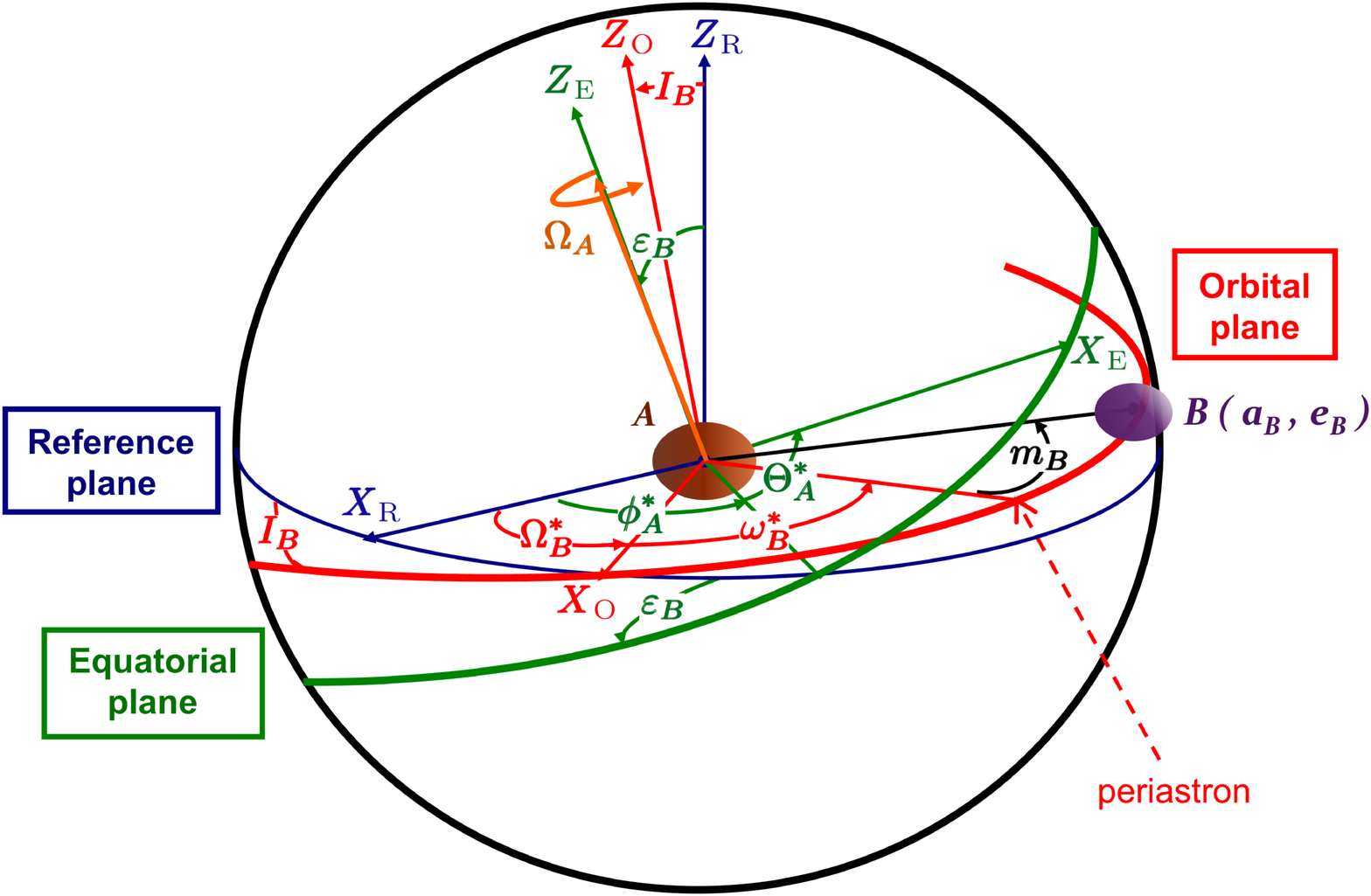}
 \caption{Inertial reference ($\mathcal{R}_R$), orbital ($\mathcal{R}_O$), and equatorial ($\mathcal{R}_E$) rotating frames, and associated Euler’s angles of orientation.
 }
 \label{fig:Reperes}
\end{figure}

The equatorial reference frame $\mathcal{R}_E$ is defined by three other Euler angles with respect to $\mathcal{R}_R$:
\vspace{-6pt}
\begin{itemize}
	\item $\varepsilon$, the obliquity of the rotation axis of $A$;
	\item $\Theta^*$, the mean sideral angle defined by ${\Omega = {\mrm d \Theta^*}/{\mrm dt}}$;
	\item $\phi^*$, the general precession angle.
\end{itemize}
\vspace{-6pt}
Refer to Figure \ref{fig:Reperes} for an illustration of the relative position of these three reference frames and the associated angles.
For convenience, all the following developpements will be done in the spin equatorial frame $\mathcal{R}_E$ of $A$ (as it has been done in MLP09).

\vspace{0.5cm}

All following results are derived from the Kaula's transform (Kaula 1962), used to explicitly express the whole generic multipole expansion in spherical harmonics of the perturbing potential $U$ in terms of the Keplerian orbital elements of $B$ in the equatorial $A$-frame:

\begin{multline}
\label{Kaula}
	\frac{1}{r_B^{l+1}} \, P_{l}^{m}(\cos\theta_B) \, e^{im\varphi_B} 
	= \frac{1}{a^{l+1}} \sum_{j=-l}^{l} \sum_{p=0}^{l} \sum_{q\in\mathbb{Z}} 
	  \left\lbrace
	\sqrt{ \frac{2l+1}{4\pi}\,\frac{(l-|j|)!}{(l+|j|)!} } \right. \\
	\left. \times d_{j,m}^{l} (\varepsilon) \, F_{l,j,p}(I) \, G_{l,p,q}(e) \, e^{\pmb i\Psi_{l,m,j,p,q}} \vphantom{\sqrt{ \frac{2l+1}{4\pi}\,\frac{(l-|j|)!}{(l+|j|)!} }} 
	\right\rbrace
\end{multline}
where $\theta_B$ and $\varphi_B$ are respectively the colatitude and the longitude of the point mass perturber $B$, and where the phase argument is given by:
\begin{equation}
\label{Psi}
	\Psi_{l,m,j,p,q}(t) = \sigma_{l,m,p,q} (n,\Omega) \, t + \tau_{l,m,j,p,q}\left( \omega^*, \Omega^*, \phi^* \right) \:.
\end{equation}
We have defined here the tidal frequency:
\begin{equation}
	\sigma_{l,m,p,q} (n,\Omega) = (l-2p+q) \, n - m \Omega \:,
\end{equation}
and the phase $\tau_{l,m,j,p,q}$:
\begin{equation}
	\tau_{l,m,j,p,q} = (l-2p)\omega^* + j(\Omega^*-\phi^*) + (l-m)\frac{\pi}{2}.
\end{equation}

We study here binary systems close enough for the tidal interaction to play a role, but we also consider that the companion is far (or small) enough to be treated as a point mass.
We then are allowed to assume the {\it quadrupolar approximation}, where we only keep the first mode of the potential, $l=2$:

\begin{multline}
\label{Ug}
U(r, \theta, \varphi, t) = \Re \left\lbrace \sum_{m=-2}^{2} \sum_{j=-2}^{2} \sum_{p=0}^{2} \sum_{q \in \mathbb{Z}} \right. \\
	\left. U_{m,j,p,q}(r) \, P_2^m(\cos\theta) \, e^{\pmb i \, \Phi_{2,m,j,p,q}(\varphi,t)} \vphantom{\sum_{m=-2}^{2}} \right\rbrace
\end{multline}
where 
\begin{equation}
	\Phi_{2,m,j,p,q}(\varphi,t) = m\varphi + \Psi_{2,m,j,p,q}(t)
\end{equation}

The functions ${U_{m,j,p,q}\left(r,\theta\right)}$ may be expressed in terms of the Keplerian elements (the semi-major axis $a$ of the orbit, its eccentricity $e$ and its inclination $I$) and the obliquity $\varepsilon$ of the rotation axis of $A$, as:
\begin{multline}
	U_{m,j,p,q}(r)= (-1)^m \, \sqrt{ \frac{(2-m)! \, (2-|j|)!}{(2+m)! \, (2+|j|)!} }  \\
	\times \frac{ \mathcal{G} \, M_B}{a^3} \,\left[ d_{j,m}^{2}(\varepsilon) \, F_{2,j,p}(I) \, G_{2,p,q}(e) \right] \, r^2 \:,
\end{multline}
where $\mathcal G$ is the gravitational constant.

The obliquity function ${d_{j,m}^{2}(\gamma)}$ is defined, for $j\se m$, by:
\begin{multline}
\label{obfunc}
	d_{j,m}^{2}(\gamma) = (-1)^{j-m} \left[\frac{ (2+j)! (2-j)! }{ (2+m)! (2-m)! }\right]^{1\over2} \\
	\times \left[ \cos\left(\frac{\gamma}{2}\right) \right]^{j+m} \, \left[ \sin\left(\frac{\gamma}{2}\right) \right]^{j-m} \, \mrm P_{2-j}^{(j-m , j+m)} (\cos \gamma) \,,
\end{multline}
where ${\mrm P_l^{\alpha,\beta} (x)}$ are the Jacobi polynomials (cf. MLP09).
The values of these functions, for indices $j < m$, are deduced from ${d_{j,m}^{2}(\pi+\gamma) = (-1)^{2-j} d_{-j,m}^{2}(\gamma)}$ or from their symmetry properties: ${d_{j,m}^{2}(\gamma) = (-1)^{j-m} d_{-j,-m}^{2}(\gamma) = d_{m,j}^{2}(-\gamma)}$; moreover, we have: ${d_{j,m}^{2}(0) = \delta_{j,m}}$.
Values are given in Table~\ref{d2jm}.

\begin{table}[!Htb]
\centering
\begin{tabular}{ c c | l}
\hline
 \hline
  $j$ & $m$& $d^{2}_{j,m}\left(\varepsilon\right)$\\
\hline
 2 & 2 & $\left(\cos\frac{\varepsilon}{2}\right)^{4}$\\
 2 & 1 & $-2\left(\cos\frac{\varepsilon}{2}\right)^{3}\left(\sin\frac{\varepsilon}{2}\right)$\\
 2 & 0 & $\sqrt{6}\left(\cos\frac{\varepsilon}{2}\right)^{2}\left(\sin\frac{\varepsilon}{2}\right)^{2}$\\
 1 & 1 & $\left(\cos\frac{\varepsilon}{2}\right)^{4}-3\left(\cos\frac{\varepsilon}{2}\right)^{2}\left(\sin\frac{\varepsilon}{2}\right)^{2}$\\
 1 & 0 & $-\sqrt{6}\cos\varepsilon\left(\cos\frac{\varepsilon}{2}\right)\left(\sin\frac{\varepsilon}{2}\right)$\\
 0 & 0 & $1-6\left(\cos\frac{\varepsilon}{2}\right)^{2}\left(\sin\frac{\varepsilon}{2}\right)^{2}$\\
\hline
\end{tabular}
\caption{\label{d2jm}Values of the obliquity function ${d_{j,m}^{2}\left(\varepsilon\right)}$ in the case where $j\se m$ obtained from Eq.~(\ref{obfunc}), (from MLP09).}
\end{table}

We also define, the inclination function $F_{2,j,p}(I)$:
\begin{multline}
	F_{2,j,p}(I) = (-1)^j \, \frac{(2+j)!}{ 4\, p! \, (2-p)! } \\
	\times \left[ \cos\left(\frac{I}{2}\right) \right]^{j-2p+2} \, \left[ \sin\left(\frac{I}{2}\right) \right]^{j+2p-2} \\
	\times \mrm P_{2-j}^{(j+2p-2 , j-2p+2)} (\cos I) \,,
\end{multline}
with the symmetry property:
\begin{equation}
\label{Fsym}
	F_{2,-j,p}(I) = \left[ (-1)^{2-j} \frac{\left(2-j\right)!}{\left(2+j\right)!}\right]F_{2,j,p}\left(I\right).
\end{equation}
Values are given in Table~\ref{F2jp}.

\begin{table}[!htb]
\centering
\begin{tabular}{c c | l}
\hline
\hline
 $j$ & $p$ & $F_{2,j,p}\left(I\right)$ \\
\hline
 0 & 0 & $\frac{3}{8}\sin^{2}I$\\
 0 & 1 & $-\frac{3}{4}\sin^{2}I+\frac{1}{2}$\\
 0 & 2 & $\frac{3}{8}\sin^{2}I$\\
 1 & 0 & $\frac{3}{4}\sin I\left(1+\cos I\right)$\\
 1 & 1 & $-\frac{3}{2}\sin I\cos I$\\
 1 & 2 & $-\frac{3}{4}\sin I\left(1-\cos I\right)$\\
 2 & 0 & $\frac{3}{4}\left(1+\cos I\right)^{2}$\\
 2 & 1 & $\frac{3}{2}\sin^{2}I$\\
 2 & 2 & $\frac{3}{4}\left(1-\cos I\right)^{2}$\\
\hline
\end{tabular}
\caption{\label{F2jp}Values of the inclination function $F_{2,j,p}(I)$. Values for indices $j < 0$ can be deduced from Eq.~(\ref{Fsym}), (from MLP09).}
\end{table}

The eccentricity functions $G_{2,p,q}(e)$ are polynomial functions having $e^q$ for argument (see Kaula, 1962). Their values for the usual sets $\{2,p,q\}$ are given in Table~\ref{G2pqe}, knowing that in the case of weakly eccentric orbits, the summation over a small number of values of $q$ is sufficient ($\intervalle{q}{-2}{2}$).
In the following, let us denote by ${\mathbb{I} = \interval{-2}{2} \times \interval{-2}{2} \times \interval{0}{2} \times \mathbb{Z}}$ the set in which the quadruple $\{m,j,p,q\}$ takes its values.

\begin{table}[!htb]
\centering
\begin{tabular}{c c | c c G l }
\hline
\hline
 $p$ & $q$ & $p$ & $q$ & $G_{2,p,q}(e)$ \\
\hline
 0 & -2 & 2 &  2 & 0\\
 0 & -1 & 2 & 1 & $-\frac{1}{2}e+\cdot\cdot\cdot$\\
 0 & 0 & 2 &  0 & $1-\frac{5}{2}e^{2}+\cdot\cdot\cdot$\\
 0 & 1 & 2 & -1 & $\frac{7}{2}e+\cdot\cdot\cdot$\\
 0 & 2 & 2 & -2 & $\frac{17}{2}e^{2}+\cdot\cdot\cdot$\\
 1 & -2 & 1 & 2 & $\frac{9}{4}e^{2}+\cdot\cdot\cdot$\\
 1 & -1 & 1 & 1 & $\frac{3}{2}e+\cdot\cdot\cdot$\\
 & & 1 & 0 & $\left(1-e^{2}\right)^{-3/2}$\\
\hline
\end{tabular}
\caption{\label{G2pqe}Values of the eccentricity function $G_{2,p,q}(e)$, (from MLP09).}
\end{table}

If we simplify the expansion of the potential in the case where spins are aligned and perpendicular to the orbital plan, where obliquity $\varepsilon$ and orbital inclination $I$ are zero, Eq.~\eqref{Ug} reduces to the expression of the potential given by Zahn (1977).

The tidal force induces a displacement of each particule constituting the planet, thus causing some deformations.
In particular, the core's surface is deformed as described by the Love theory (Love 1911).

\subsection{Dynamical equations for a solid body and their boundary conditions}
\label{subsection:dyn_eq}

To describe the internal evolution of the main component $A$ submitted to the perturbations induced by the tidal potential presented above, we use the Eulerian formalism (Dahlen et al. 1999).
The system of equations, needed to follow the motion of a particule, is composed by the Eulerian momentum \eqref{eq_s} and mass \eqref{eq_rho} conservation laws, and the Poisson equation \eqref{eq_Phi} satisfied by the potential $\Phi$ of self-gravitation:  
\begin{subequations}
\label{syst_elastic}
\begin{flalign}
  &\rho \, \frac{\partial^2 \bf s}{\partial t^2} = \nab \cdot \tenseur{\sigma} + \rho \, \nab \left( \Phi + U \right) \:, \label{eq_s}\\ 
  &\frac{\partial \rho}{\partial t} + \nab \cdot \left( \rho \, \frac{\partial \mbf s}{\partial t} \right) = 0 \:, \label{eq_rho}\\ 
  &\nab^2 \Phi = - 4 \pi \mathcal G \rho \:, \label{eq_Phi}
\end{flalign}
where $\bf s$ designates the displacement vector and $\tenseur{\sigma}$ is the stress tensor.
We complete this system with the constitutive equation to link the stress exerted on the body to the resulting deformation.
Assuming that tidal effect corresponds to a traction applied on the body, with no rotational contribution, the deformation tensor reduces to the strain tensor $\tenseur{\epsilon}$:
\begin{equation}
	\tenseur{\epsilon} =  \frac{1}{2} \left[ \nab \mbf s + \left( \nab \mbf s \right)^{\mrm T} \right] \:,
\end{equation}
where $\tenseur{h}^{\mrm T}$ designates the transposed tensor of $\tenseur{h}$.
We then get a relation linking the stress tensor $\tenseur{\sigma}$ to the strain tensor $\tenseur{\epsilon}$ that accounts for the rheology of the body, and that we represent by a function ${\mathcal F_\mrm{rh}}$:
\begin{equation}
\label{eq_const}
	\tenseur{\sigma} = {\mathcal F_\mrm{rh}} (\tenseur{\epsilon}) \:.
\end{equation}
\end{subequations}

To solve this system \eqref{syst_elastic}, we need to apply boundary conditions to the five previous equations, assuming that there is no displacement \eqref{CLs} neither attraction \eqref{CLPot} at the center of mass $r=0$, the gravitational potential has to be continuous \eqref{CLPhi} and the Lagrangian traction has to vanish \eqref{CLT} at the surface $r=R$:
\begin{subequations}
\label{CL_elastic}
\begin{flalign}
	&\left. \mbf s \right|_{r=0}= \mbf 0 \:, \label{CLs} \\
	&\left. \left( \Phi + U \right) \right|_{r=0} = \mbf 0 \:, \label{CLPot} \\
	&\left[ \Phi \right]_{R^{-}}^{R^{+}} = 0 \quad,\quad\text{i.e. : }\quad \left[ \frac{\partial \Phi}{\partial r} + 4\pi \, \mathcal G \, \rho \, s_r \right]_{R^{-}}^{R^{+}} = 0 \:,  \label{CLPhi}\\
	&\left. \left( {\mbf e_r} \cdot \tenseur{\sigma} \right) \right|_ {r=R} = 0 \label{CLT}\:.
\end{flalign}
\end{subequations}

\subsection{Linearization of the system}
\label{subsection:dyn_eq_lin}

Assuming that tidal effects, and thus the resulting elastic deformation, are small amplitude perturbations to the hydrostatic equilbrium, we are allowed to linearize the system \eqref{syst_elastic} and its boundary conditions \eqref{CL_elastic}.
To do so, we expand a scalar quantity $X$ as:
\begin{equation}
	X \left(r,\theta,\varphi,t\right) = X_0(r) + X' \left(r,\theta,\varphi,t\right) \, ;
\end{equation}
$X_0$ designates the spherically symmetrical profile of $X$, and $X'$ represents the perturbation due to the tidal potential. 
The displacement $\mbf s$ and the tidal potential $U$ are also considered as perturbations.
Thus, correct to first order in $|| \mbf s ||$, we obtain the following form of system \eqref{syst_elastic}:
\begin{subequations}
\label{syst_elastic_lin}
\begin{flalign}
	\rho_0 \, \frac{\partial^2 \bf s}{\partial t^2} &= \nab \cdot \tenseur{\sigma} + \rho_0 \, \nab \left( \Phi' + U \right) + \rho' \, \nab \Phi_0 \:, \label{eq_s_lin}\\ 
  	\rho' + \nab \cdot \left( \rho_0 \, \bf s \right) &= 0 \:, \label{eq_rho_lin}\\ 
	\nab^2 \Phi' &= - 4 \pi \mathcal G \rho' \:, \label{eq_Phi_lin} \\
	\tenseur{\sigma} &= \left( K - \frac{2}{3} \mu \right) \mrm{tr} \left( \tenseur{\epsilon} \right) \, \tenseur{I} + 2 \mu \tenseur{\epsilon} \:, \label{eq_const_lin}
\end{flalign}
\end{subequations}
where we made use of the Hooke's law \eqref{eq_const_lin}, which is a linear constitutive law that governs elastic materials as long as the load does not exceed the material's elastic limit, in the case of an isotropic material (i.e.: whose properties are independent of direction in space). 
It means that strain is directly proportional to stress, through the bulk modulus $K$ and the shear modulus $\mu$.
The reference state is drawn from an up-to-date planetary structure model.
It is governed by the following Poisson equation and the static momentum equation:
\begin{subequations}
\begin{flalign}
	\nab^2 \Phi_0 &= - 4 \pi \mathcal G \rho_0 \:, \label{eq_Phi0} \\
	\nab P_0 &= \rho_0 \, \nab \Phi_0  \:,
\end{flalign} 
\end{subequations}
where we made use of the following convention for the gravity: ${\mbf g_0 = \nab \Phi_0}$.

\subsection{Analytical solutions for an homogeneous incompressible body}
\label{subsection:solutions}

To solve the linear system \eqref{syst_elastic_lin}, we expand all scalar quantities in spherical harmonics $Y_l^m(\theta, \varphi)$.
Moreover, as all vectorial quantities that intervene in Eqs.~(\ref{eq_s_lin}-\ref{eq_rho_lin}) are poloidal, we may expand them in the basis of vectorial spherical harmonics ${\left[ \mbf R_l^m(\theta, \varphi) , \mbf S_l^m(\theta, \varphi) \right]}$, where $\mbf R$ refers to the radial part and $\mbf S$ to the spheroidal part of a given vector (Rieutord 1987, Mathis \& Zahn 2005):
\begin{subequations}
\begin{empheq}
[left=\displaystyle{ \forall (l,m) \in \mathbb{N}\times \textrm{\textlbrackdbl}-l,l\textrm{\textrbrackdbl}\:,\:\: \empheqlbrace \:\:}] {flalign}
	\mbf R_l^m(\theta, \varphi) &= Y_l^m(\theta, \varphi) \, \mbf e_r \:, \\
	\mbf S_l^m(\theta, \varphi) &= \nab_{\mrm S} \left[ Y_l^m(\theta, \varphi) \right] \:, &
\end{empheq}
\end{subequations}
where $\nab_{\mrm S}$ designates the horizontal gradient:
\begin{equation}
	\nab_{\mrm S} = \frac{\partial \, \cdot}{\partial \theta} \, \mbf e_{\theta} + \frac{1}{\sin \theta} \frac{\partial \, \cdot}{\partial \varphi} \, \mbf e_{\varphi} \:.
\end{equation}

We introduce six radial functions $y_{\{1...6\}}^m(r)$ (Takeuchi \& Saito 1972) to expand all quantities in spherical harmonics, at the quadrupolar approximation ($l=2$):
\begin{subequations}
\label{syst:spheric_exp}
\begin{itemize}
	\item the displacement: 
		\begin{multline}
			\mbf s (r,\theta,\varphi,t) = \sum_{(m,j,p,q) \in \mathbb{I}} \left[ 
			y_1^m(r) \, \mbf R_2^m(\theta, \varphi) \right. \\
			\left. + y_3^m(r) \, \mbf S_2^m(\theta, \varphi) \right] \, e^{\pmb i\Psi_{2,m,j,p,q}(t)} \:,
		\end{multline}
	\item the total potential:
		\begin{equation}		
			(U+\Phi') (r,\theta,\varphi,t) = \sum_{(m,j,p,q) \in \mathbb{I}} y_5^m(r) \, Y_2^m(\theta, \varphi) \, e^{\pmb i\Psi_{2,m,j,p,q}(t)} \:,
		\end{equation}
	\item the Lagrangian traction:
		\begin{multline}
			\mbf e_r \cdot \, \tenseur{\sigma}(r,\theta,\varphi,t) = \sum_{(m,j,p,q) \in \mathbb{I}} \left[ 
			y_2^m(r) \, \mbf R_2^m(\theta, \varphi) \right. \\
			\left. + y_4^m(r) \, \mbf S_2^m(\theta, \varphi) \right] \, e^{\pmb i\Psi_{2,m,j,p,q}(t)} \:,
		\end{multline}
	\item the Lagrangian attraction (introduced to express the continuity of the gradient of the potential):
		\begin{equation}
		 	\forall \intervalle{m}{-2}{2}, \: 
		 	y_6^m(r) =  \frac{\mrm d}{\mrm d r} \left[ y_5^m(r) \right] - 4\pi \mathcal G \rho_0 y_1^m(r) \:.
		\end{equation}
\end{itemize}
\end{subequations}

The linear system governing the radial functions $y_{\{1...6\}}^m(r)$ is given in Appendix.
In the case of an incompressible (${K \rightarrow +\infty}$) and homogeneous body (${\mu, \, \rho_0 = \mit{cst}}$), the system \eqref{syst_elastic_lin} constrained by boundary conditions \eqref{CL_elastic} admits the following solutions, considering the expansion \eqref{syst:spheric_exp}: $\displaystyle \forall \intervalle{m}{-2}{2} $,
\begin{subequations}
\label{syst:solutions}
\begin{flalign}
	y_1^m(r) &= \sum_{j,p,q} \, \frac{k_2}{r\,R\,g_s} \, \left( \frac{8}{3} R^2 - r^2 \right) \, U_{m,j,p,q}(r) \:, \\
	y_2^m(r) &= \sum_{j,p,q} \, \left[ 2\mu \dfrac{k_2}{r^2\,R\,g_s} \left( \dfrac{8}{3} R^2 + \dfrac{1}{2} r^2 \right) \right. \nonumber\\
	 	&\qquad\quad + \dfrac{4}{3} \pi \mcal G \rho^2 \dfrac{k_2}{R\,g_s} \left( \dfrac{8}{3} R^2 - \dfrac{1}{2} r^2 \right) 
	 	  \nonumber\\
	 	&\qquad\qquad\qquad  - \rho (1+k_2) \Bigg]  U_{m,j,p,q}(r) \:,  \\ \nonumber\\
	y_3^m(r) &= \sum_{j,p,q} \, \frac{k_2}{r\,R\,g_s} \, \left( \frac{4}{3} R^2 - \frac{5}{6} r^2 \right) \, 
				U_{m,j,p,q}(r)\:, \\
	y_4^m(r) &= \sum_{j,p,q} \frac{8\mu}{3} \, \frac{k_2}{r\,R\,g_s} \, (R^2 - r^2) \, U_{m,j,p,q}(r) \:, \\
	y_5^m(r) &= \sum_{j,p,q} \, (1 + k_2) \, U_{m,j,p,q}(r) \:, \\
	y_6^m(r) &= \sum_{j,p,q} \, \left[ \frac{2 (1+k_2)}{r}  \right. \nonumber\\
			& \left. + 4\pi \mcal G \rho_0 \, \frac{k_2}{r\,R\,g_s} \, \left( r^2 - \frac{8}{3} R^2 \right) \right] U_{m,j,p,q}(r) \:,
\end{flalign}
\end{subequations}
where we have introduced the acceleration of gravity at the surface $g_s$ and the second-order Love number $k_2$.
The latter compares the perturbed part $\Phi'(R)$ of the self-gravitational potential at the surface of a fully-solid planet of mean radius $R$, deformed by tidal force, with the tidal perturbing potential $U(R)$:
\begin{equation}
\label{k2def}
	k_2  \stackrel{\text{def}}{=}  \frac{\Phi'(R)}{U(R)} \: .
\end{equation}
The expression of $k_2$ is established in Sect.~\ref{subsect_fluid_env}, for an ocean-free planet (Eq.~\ref{k2}) or a two-layer planet (Eq.~\ref{k2'}).

\begin{figure*}[!htb]
      \centering
      \includegraphics[width=0.3\linewidth]{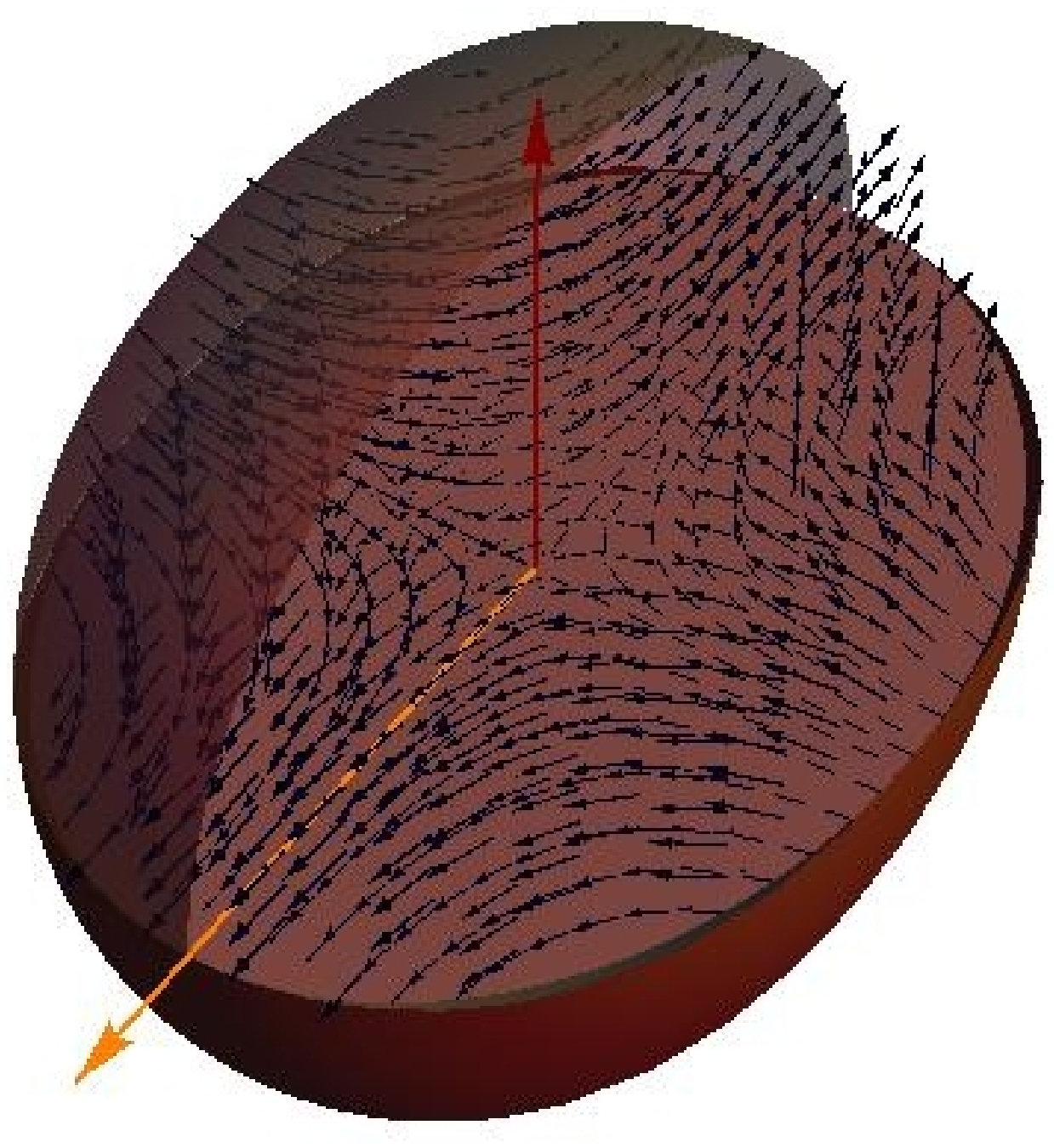}
      \includegraphics[width=0.38\linewidth]{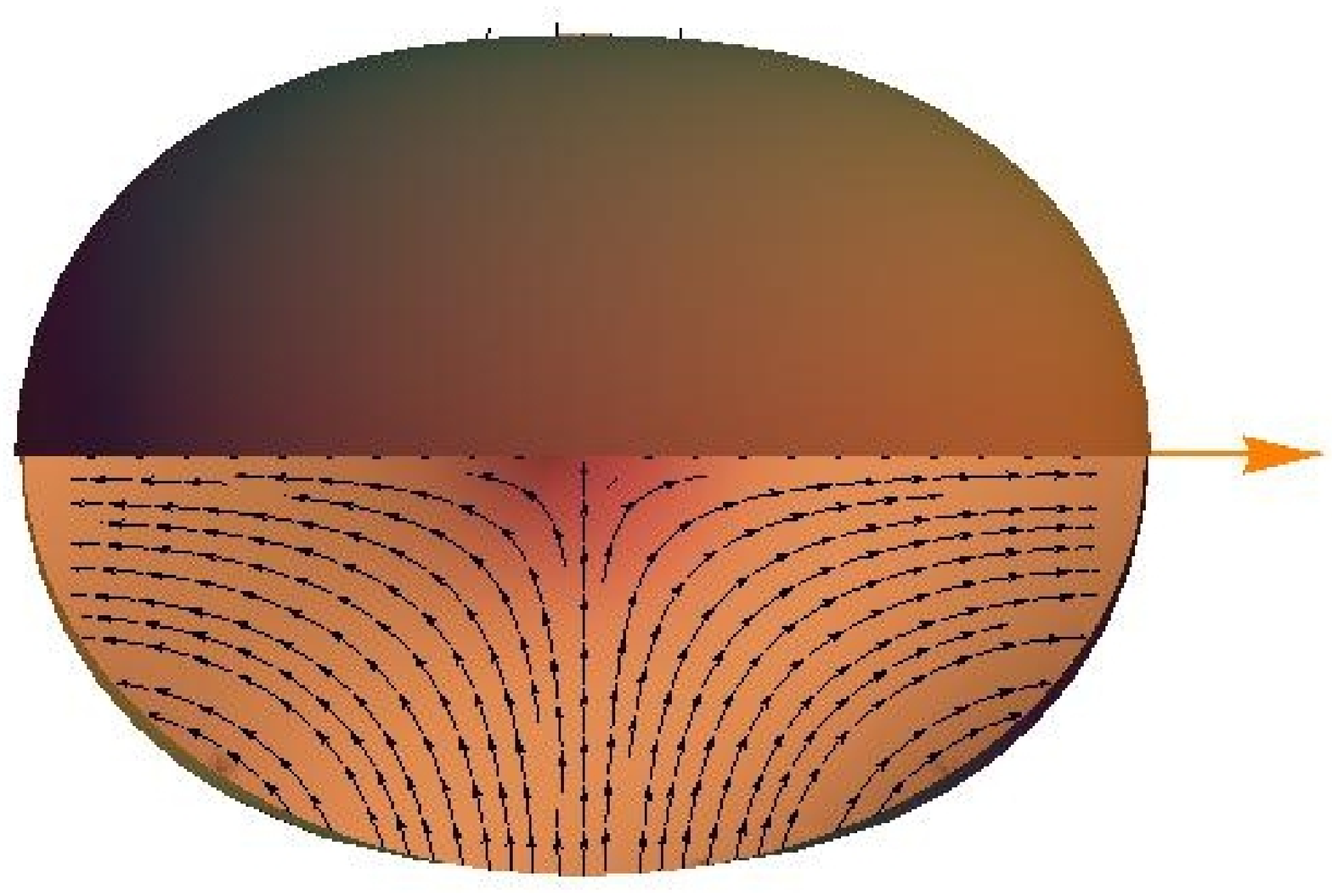}
      \includegraphics[width=0.30\linewidth]{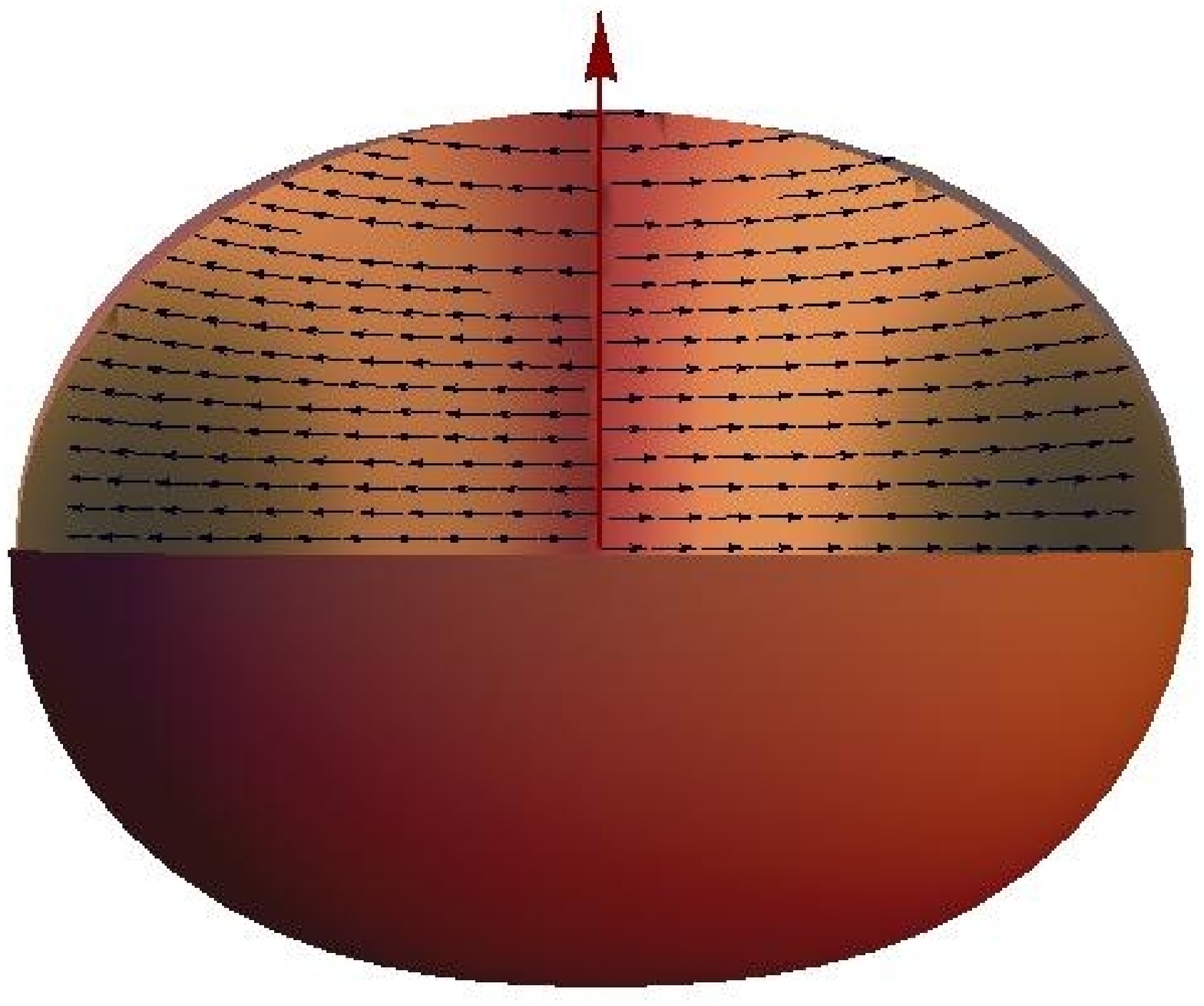}
      \caption{{\it Left}: tidal displacement $\mbf s$. {\it Middle}: equatorial slice of $\mbf s$. {\it Right}: meridional slice of $\mbf s$. The orange arrow indicates the direction of the perturber $B$, the red one corresponds to the rotation axis of $A$. The two slices are planes of symmetry.}
\end{figure*}


\section{Modified elastic tidal theory in presence of a fluid envelope}
\label{subsect_fluid_env}
We now assume that planet $A$ is not entirely solid, but has a fluid envelope.
We follow the method proposed by Dermott (1979)
to evaluate how the anelastic dissipation is modified by the presence of a fluid layer surrounding the solid region.
The first step consists in determining the behaviour of the elastic response in this configuration.
We denote by $R_c$ (resp. $R_p$) the mean radius of the solid core (resp. of the whole planet, including the height of the fluid layer); $\rho_c$ and $\rho_o$ designate the density of the core and the ocean, both assumed to be uniform, as a first step.
More generally, all quantities will be written with a ``$c$" subscript when evaluated at the core boundary, and with
a ``$p$" subscript if taken at the surface of the planet.
The evolution of the system is described in the orbital frame ${\mcal R_O: \{A, X_O, Y_O, Z_O\}}$ centered on $A$ and comoving with the perturber $B$.
We will use polar coordinates $(r,\Theta)$ to locate a point $P$, where $r$ is the distance to the center of $A$, and $\Theta$ is the angle formed by the radial vector and the line of centers.

\subsection{Vertical deformation at the boundary of the core}

In $\mcal R_O$, the tidal potential takes the form (Dermott 1979)
\begin{equation}
\label{U2bis}
	U(\mbf r) = - \zeta(r) \, g(r) \, P_2(\cos \Theta) = - \zeta_c \, g_c \, \frac{r^2}{R_c^2} \, P_2(\cos \Theta) \:,
\end{equation}
where we have introduced the tidal height
\begin{equation}
\label{zeta_r}
	\zeta(r) = \frac{M_B}{M(r)} \, \left( \frac{r}{a} \right)^3 \, r \:,
\end{equation}
and the gravity
\begin{equation}
\label{g_r}
	g(r) = \frac{\mcal{G} M(r)}{r^2} \:,
\end{equation}
$M(r)$ being the fraction of mass of the planet inside the radius $r$.

The expression of the tidal potential in the rotating frame of $B$ (Eq.~\ref{U2bis}) is linked to its expression in the equatorial inertial frame (Eq.~\ref{Ug}), through the Kaula's transform (Eq.~\ref{Kaula}).
Indeed, the Legendre polynomia summation formula 
\begin{multline}
\label{P2costheta}
	P_2(\cos \Theta) = 
	\Re \left[ \sum_{m=0}^2 \frac{(2-m)!}{(2+m)!} (2-\delta_{0,m}) \, P_2^m(\cos\theta) \, {\mbf e}^{\pmb i m\varphi} \right. \\ 
	\times \left. P_2^m(\cos\theta_B) \, {\mbf e}^{- \pmb i m \varphi_B} \vphantom{\sum_n} \right]
\end{multline}
involves the term ${P_2^m(\cos\theta_B) \, {\mbf e}^{- \pmb i m \varphi_B}}$ in Eq.~\eqref{U2bis} that has to be transformed following Eq.~(\ref{Kaula}-\ref{Psi}) to obtain Eq.~\eqref{Ug}.\\

In this section we are interested in the mofication of Love numbers due to the presence of a fluid envelope on top of the solid core.
Thus, we will focus on the deformations of the core's surface, and particularly on the vertical displacements.
Love (1911) proved that tidal deformations could be described by the same harmonic function than the tidal potential which causes it.
Therefore the equations of the core and planet boundaries are respectively of the form
\begin{subequations}
\label{eqs_surf}
\begin{flalign}
	r_c \equiv R_c + s_r(R_c) &= R_c \left[ 1 + S_2 \, P_2(\cos\Theta) \right] \:, \label{eq_surf_core} \\
	r_p                       &= R_p \left[ 1 + T_2 \, P_2(\cos\Theta) \right] \:. \label{eq_surf_planet}
\end{flalign}
\end{subequations}

Thus, ${s_r(R_c) = R_c \, S_2 \, P_2(\cos\Theta)}$ represents the radial displacement at the core's boundary corresponding to the vertical tidal deformation of amplitude ${S_2 \, P_2(\cos\Theta)}$.
In 1909, Love defined the number $h_2$, as the ratio between the amplitude of the vertical displacement at the surface of the planet and the equilibrium tidal height (disturbing potential divided by undisturbed surface gravity, both taken at the surface of the core) in the case of a fully-solid planet.
Solving the whole system of equations, he determined its expression as
\begin{equation}
\label{h2}
	h_2 \stackrel{\text{def}}{=} \frac{s_r(R_c)}{U(R_c) / g_c} \equiv \frac{R_c \, S_2}{\zeta_c} = \frac{5}{2} \, \frac{1}{1+\bar{\mu}} \: ,
\end{equation}
where $\bar{\mu}$ is called the effective rigidity, in the sense that it evaluates the relative importance of elastic and gravitational forces:
\begin{equation}
\label{mu_bar}
	\bar{\mu} = \frac{19 \mu}{2 \rho_c g R_c} \: .
\end{equation}
In presence of the fluid envelope, the ratio between the amplitude of the tidal surface vertical displacement and the tidal height will be modulated by a multiplicative factor $F$, due to the additional loading exerted by the tidally-deformed fluid layer.
We may then introduce a new notation $h^F_2$ for the modified Love number in presence of a fluid envelope:
\begin{equation}
\label{h2'}
	h^F_2 \stackrel{\text{def}}{=} \frac{s_r(R_c)}{U(R_c)/g_c} \equiv \frac{R_c \, S_2}{\zeta_c} = F \, h_2  = F \times \frac{5/2}{1+\bar{\mu}} \: .
\end{equation}
We have now to express this factor in function of the parameters of the system.
To do so, we have to list all the forces acting on the surface of the core.
Before carrying out the study of these forces, let us introduce a specific notation.
All physical quantities $X(\mbf r)$ will be separated in two terms: the first corresponds to the constant part that does not depend on where the quantity is calculated; the second one (called the "effective deforming" contribution and denoted $X'(\mbf r)$) is a term proportional to the spherical surface harmonic $P_2$ (see \ref{P2costheta}).

\subsection{Gravitational forces acting on the surface of the core}

The planet is not only subjected to the direct action of the tidal potential $U$, but also to the self-gravitational potential $\Phi$ perturbed by the first.
In calculating the latter, we have to consider both contributions of the solid core and the fluid envelope, $\Phi_c$ and $\Phi_o$ respectively.

At any point $\mbf r$ of the core, $\Phi_c (\mbf r)$ corresponds to the internal potential created by the core :
\begin{equation}
\label{Phic1}
	\Phi_c (\mbf r) = - \frac{g_c}{R_c} \left( \frac{3 R_c^2 - r^2}{2} + \frac{3}{5} \, r^2 \, S_2 \, P_2 \right) \: .
\end{equation}
At the same point $\mbf r$, $\Phi_o (\mbf r)$ is the internal potential created by the fluid shell of density $\rho_o$ and of lower and upper surface boundaries $r_c$ and $r_p$ respectively:
\begin{multline}
	\Phi_o (\mbf r) = - \frac{\rho_o}{\rho_c} \, \frac{g_c}{R_c} \, \left[ 
		\left( \frac{3 R_p^2 - r^2}{2} + \frac{3}{5} \, r^2 \, T_2 \, P_2 \right)  \right.\\
		\left. - \left(  \frac{3 R_c^2 - r^2}{2} + \frac{3}{5} \, r^2 \, S_2 \, P_2  \right) 
	\right] \: .
\end{multline}

Therefore ${V(\mbf r) = U(\mbf r) + \Phi_c(\mbf r) + \Phi_o(\mbf r)}$ has the following expression, at any point $\mbf r$ inside the core:
\begin{subequations}
\label{Pot_tot_c}
\begin{equation}
\label{Vc}
	V(\mbf r) = - \frac{1}{2} \, \frac{g_c}{R_c} \, \left[ -r^2 + 3 R_c^2 \left( 1 - \frac{\rho_o}{\rho_c} +  \frac{\rho_o}{\rho_c} \, \frac{R_p^2}{R_c^2} \right) \right] - V'(\mbf r) \: ,
\end{equation}
where the effective deforming potential is expressed by
\begin{equation}
\label{Vc'}
	V'(\mbf r) = - Z \, r^2 \, P_2 \: ,
\end{equation}
$Z$ being a constant that depends on the characteristics of the planet:
\begin{equation}
\label{Z}
 Z =  \frac{g_c}{R_c} \, \left[ \frac{\zeta_c}{R_c} + \frac{3}{5} \, \frac{\rho_o}{\rho_c} (T_2-S_2) + \frac{3}{5} S_2 \right] \: .
\end{equation}
\end{subequations}

We then obtain its expression, correct to first order in $S_2 P_2$ or $T_2 P_2$, at any point ${\mbf{r_c} = r_c \, \mbf{e_r}}$ of the surface of the core:
\begin{subequations}
\label{Pot_tot_sc}
\begin{equation}
\label{Vc}
	V(\mbf{r_c}) = - g_c \, R_c \, \left[ \left( 1 - \frac{3}{2} \frac{\rho_o}{\rho_c} \right) + \frac{3}{2} \frac{\rho_o}{\rho_c} \, \frac{R_p^2}{R_c^2} \right] + V'(\mbf{r_c}) \: ,
\end{equation}
where 
\begin{equation}
\label{Vsc'}
	V'(\mbf{r_c}) = - Z_c \, R_c^2 \, P_2 \: ,
\end{equation}
$Z_c$ being a constant that depends on the characteristics of the planet:
\begin{equation}
\label{Zc}
 Z_c =  \frac{g_c}{R_c} \, \left[ \frac{\zeta_c}{R_c} + \frac{3}{5} \, \frac{\rho_o}{\rho_c} (T_2-S_2) - \frac{2}{5} S_2 \right] \: .
\end{equation}
\end{subequations}

Chree (1896) showed that the deformation of the core's surface under the gravitational forces (that derive from the effective deforming potential $V'$) is the same as that which would result from the outward normal traction $f^{T_N}_1$ applied at its mean surface $r=R_c$:
\begin{equation}
\label{TN1}
	f^{T_N}_1(R_c) = \rho_c \, Z \, R_c^2 \, P_2 \: .
\end{equation}

\subsection{Total effective normal traction acting on the surface of the core}

The mean surface of the core is subjected to two additional forces induced by both the loading of the core and the loading of the ocean, tidally deformed.

First, the pressure due to the differential overloading of the deformed elastoviscous matter on the mean surface of radius $R_c$ is given by the product of the local gravity $g_c$, the density of the core $\rho_c$ and the solid tidal height $R_c S_2$:
\begin{equation}
\label{TN2}
	f^{T_N}_2(R_c) = - \rho_c \, g_c \, R_c \, S_2 \, P_2 \: .
\end{equation}

We also have to take into account the oceanic hydrostatic pressure.
Following Zahn (1966) and Remus et al. (2012), we express all scalar quantities ${X(r,\Theta)}$ as the sum of their spherically symmetrical profile $X^0(r)$ and their perturbation ${X'(r,\Theta)}$ due to the tidal potential ${U(r,\Theta) \propto P_2(\cos\Theta)}$:
\begin{equation}
	\label{Xlin}
	X(r,\Theta) = X^0(r) + X'(r,\Theta) \equiv X^0(r) + \hat{X}(r) \, P_2(\cos\Theta)\:.
\end{equation}
The perturbations of pressure ${P'(r,\theta)}$ obey the relation of the hydrostatic equilibrium which is of the following form, correct to first order in $P_2(\cos\Theta)$:
\begin{equation}
	\label{Hydr_Eq}
	\nabla P' = \rho_o \, \nabla V' + \rho_o' \, \nabla V^0 \:.
\end{equation}
Therefore, the $\Theta$-projection of \eqref{Hydr_Eq} leads to 
\begin{equation}
	P'(r,\Theta) = \rho_o \, V'(r,\Theta) \:.
\end{equation}
Finally, since only the variable part of the pressure, {\it i.e.} $P'$, contributes to the normal effective traction $f^{T_N}_3$ that acts on the mean surface of core, this latter takes the following expression:
\begin{equation}
\label{TN3}
	f^{T_N}_3(R_c) = P'(\mbf{r_c}) = \rho_o \, V'(\mbf{r_c}) = - \rho_o \, Z_c \, R_c^2 \, P_2 \: .
\end{equation}

\begin{figure}[!htb]
      \centering
      \includegraphics[width=0.7\linewidth]{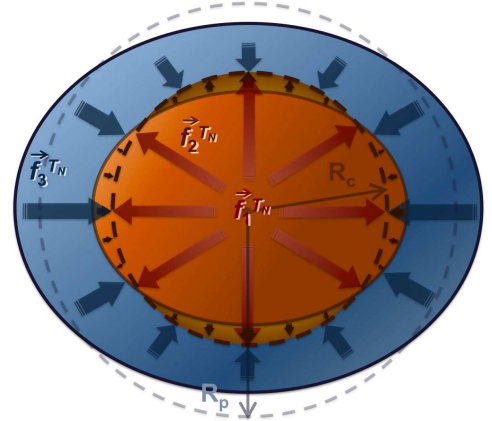}
      \caption{
      Balance of forces that act on the mean surface of the core $r=R_c$: $f^{T_N}_1 (R_c)$ are the gravitational forces, $f^{T_N}_2 (R_c)$ the loading of the solid tide and $f^{T_N}_3 (R_c)$ the hydrostatic pressure.
      \label{fig:bilan_forces}
      }
\end{figure}

The sum of these three forces, represented on Fig.~\ref{fig:bilan_forces}, corresponds to the total normal effective traction ${f^{T_N} (R_c) =  f^{T_N}_1 (R_c) + f^{T_N}_2 (R_c) + f^{T_N}_3 (R_c)}$ that deforms the mean surface of the core.
Using Eqs.~(\ref{TN1}-\ref{TN2}-\ref{TN3}), we get:
\begin{equation}
	f^{T_N} (R_c) = \left( \rho_c \, Z - \rho_o \, Z_c \right) \, R_c^2 \, P_2 - \rho_c \, g_c \, R_c \, S_2 \, P_2 \: ,
\end{equation}
where the expressions of $Z$ and $Z_c$ are given by equations \eqref{Z} and \eqref{Zc} respectively, so that
\begin{equation}
	f^{T_N} (R_c) = X \, P_2 (\cos\Theta) \: ,
\end{equation}
where we have denoted by $X$ the following quantity:
\begin{equation}
\label{X}
	X = \frac{2}{5} \, \rho_c \,  g_c \, R_c \left( 1 - \frac{\rho_o}{\rho_c} \right) \, \left[ \frac{5}{2} \frac{{\zeta}_c}{R_c} - S_2 + \frac{3}{2} \frac{\rho_o}{\rho_c} \left( T_2 - S_2 \right) \right] \: .
\end{equation}

\subsection{Amplitude of the vertical deformation}

According to Melchior (1966), a deforming potential $\mcal{U}_2$ of second order produces a deformation at each point $\mbf{r_c}$ of the surface of the core which radial component takes the form
\begin{equation}
	{\epsilon}_{rr} = \frac{\left(8 R_c^2 - 3 r_c^2\right)}{19 \mu} \frac{\rho_c \,\mcal{U}_2}{r_c^2} \: .
\end{equation}
Correct to first order in $P_2$, recalling that ${r_c = R_c \left( 1 + S_2 P_2 \right)}$, this reduces to:
\begin{equation}
	{\epsilon}_{rr} = \frac{5}{19 \mu} \, \rho_c \mcal{U}_2 \: ,
\end{equation}
where $\rho_c \mcal{U}_2$ is the deforming traction ${f^{T,N}(R_c) = X P_2}$ applied on the core.
Furthermore, since the amplitude of the displacement is also given by ${{\epsilon}_{rr} = S_2 P_2}$ (Eq.~\ref{eq_surf_core}), we have the following equality
\begin{equation}
	S_2 P_2 = \frac{5}{19 \mu} \, X P_2 \: .
\end{equation}
Therefore the relation \eqref{mu_bar} beetween $\mu$ and $\bar{\mu}$ and the expression \eqref{X} of $X$ enable us to relate the deformation of the surfaces of the core ($S_2$) and of the ocean ($T_2$) as
\begin{equation}
\label{S2}
	S_2 = \frac{1}{\bar{\mu}} \, \left( 1 - \frac{\rho_o}{\rho_c} \right) \left[ \frac{5}{2} \frac{{\zeta}_c}{R_c} - S_2 + \frac{3}{2} \frac{\rho_o}{\rho_c} (T_2 - S_2) \right] \: . 
\end{equation}

By definition, given in Eq.~\eqref{h2'}, the impedance $F$ is of the form
\begin{equation}
\label{Fg}
	F = \frac{2}{5} \, (1+\bar{\mu}) \, \frac{R_c}{{\zeta}_c} \, S_2 \: .
\end{equation}

Since the surface of the planet is an equipotential, the total potential $V$ takes a constant value at any point $\mbf{r_p}$ of the surface of the ocean:
\begin{equation}
	V(\mbf{r_p}) \equiv U(\mbf{r_p}) + \Phi(\mbf{r_p}) = \mit{cst} \: .
\end{equation}
As $V - V' = \mit{cst}$ by definition, we get the simplier condition:
\begin{equation}
\label{equipot}
	V'(\mbf{r_p}) \equiv U(\mbf{r_p}) + \Phi'(\mbf{r_p}) = \mit{cst} \: .
\end{equation} 

At a point $\mbf r$ of the ocean, $\Phi_c (\mbf r)$ corresponds to the external potential created by the core:
\begin{equation}
	\Phi_c (\mbf r) = - g_c \, R_c^2 \left( \frac{1}{r} + \frac{3}{5} \, \frac{R_c^2 \, S_2}{r^3}  \, P_2 \right) \: .
\end{equation}
At the same point $\mbf r$, $\Phi_o (\mbf r)$ is the internal potential created by the fluid shell of density $\rho_o$ and of lower and upper surface boundaries $r_c$ and $r_p$ respectively:
\begin{equation}
\begin{split}
	\Phi_o (\mbf r) = 
		& - g_c \, \frac{R_p^3}{R_c}  \, \frac{\rho_o}{\rho_c} \, \left( \frac{3 R_p^2 - r^2}{2 R_p^3} + \frac{3}{5} \, \frac{r^2 T_2}{R_p^3} \, P_2 \right) \\
		& + g_c \, R_c^2 \, \frac{\rho_o}{\rho_c} \, \left( \frac{1}{r} + \frac{3}{5} \, \frac{R_c^2 \, S_2}{r^3}  \, P_2 \right)
		\: .
\end{split}
\end{equation}

Therefore ${V(\mbf r) = U(\mbf r) + \Phi_c(\mbf r) + \Phi_o(\mbf r)}$ has the following expression, at any point $\mbf r$ inside the ocean:
\begin{subequations}
\label{Pot_tot_o}
\begin{equation}
\label{Vo}
	V(\mbf r) = - g_c \, R_c^2 \, \left[ \left( 1 - \frac{\rho_o}{\rho_c} \right) \, \frac{1}{r}  - \frac{\rho_o}{\rho_c} \frac{r^2}{2 R_c^3} + \frac{3}{2} \frac{\rho_o}{\rho_c} \frac{R_p^2}{R_c^3} \right] + V'(\mbf r) \: ,
\end{equation}
where the effective deforming potential is expressed by
\begin{equation}
\label{Vo'}
	V'(\mbf r) = - g_c \, R_c^2 \, W(r) \, P_2 \: ,
\end{equation}
$W$ being a function of the distance $r$ to the center of the planet:
\begin{equation}
\label{W}
 W(r) = \frac{{\zeta}_c}{R_c} \, \frac{r^2}{R_c^3} + \frac{3}{5} \, \frac{\rho_o}{\rho_c} \, \frac{r^2}{R_c^3} \, T_2 + \frac{3}{5} \, \left( 1 - \frac{\rho_o}{\rho_c}  \right) \, \frac{R_c^2}{r^3} \, S_2 \: .
\end{equation}
\end{subequations}

We then obtain its expression at a point ${\mbf{r_p} = r_p \, \mbf{e_r}}$ of the surface of the planet:
\begin{subequations}
\label{Pot_tot_sp}
\begin{equation}
\label{Vsp}
	V(\mbf{r_p}) =  - g_c  \left[ \left( 1 - \frac{\rho_o}{\rho_c} \right) \, \frac{R_c^2}{R_p}  + \frac{\rho_o}{\rho_c} \frac{R_p^2}{R_c} \right] + V'(\mbf{r_p}) 
	\: ,
\end{equation}
where 
\begin{equation}
\label{Vsp'}
	V'(\mbf{r_p}) = - g_c \, W_p \, P_2 \: ,
\end{equation}
$W_p$ being a constant that depends on the planet characteristics:
\begin{multline}
\label{Wp}
W_p = \frac{R_p^2}{R_c^2} \, {\zeta}_c
	+ \left[ \frac{3}{5} \, \left( 1 - \frac{\rho_o}{\rho_c} \right) \, \frac{R_c^4}{R_p^3} \right] \, S_2   \\ 
	+ \left[ - \frac{2}{5} \, \frac{\rho_o}{\rho_c} \, \frac{R_p^2}{R_c} - \left( 1 - \frac{\rho_o}{\rho_c} \right) \, \frac{R_c^2}{R_p} \right] \, T_2
	\: .
\end{multline}
\end{subequations}

The condition \eqref{equipot} takes then the form:
\begin{equation}
\label{zetaR}
	\frac{{\zeta}_c}{R_c} = 
		- \frac{3}{5} \, \left( \frac{R_c}{R_p} \right)^5 \, \left( 1 - \frac{\rho_o}{\rho_c} \right) \, S_2 
		+ \frac{2}{5} \, \frac{\rho_o}{\rho_c} \alpha \, T_2
	\, ,
\end{equation}
where 
\begin{equation}
\label{alpha}
	\alpha = 1 + \frac{5}{2} \, \frac{\rho_c}{\rho_o} \, \left( \frac{R_c}{R_p} \right)^3 \, \left( 1 - \frac{\rho_o}{\rho_c} \right) \: .
\end{equation}
We may eliminate the variable $T_2$ thanks to Eq.~\eqref{S2}:
\begin{multline}
	\frac{2}{5} \, \frac{\rho_o}{\rho_c} \alpha \, T_2 = 
		\frac{ \frac{2}{5} \, \alpha }{ \left( \alpha + \frac{3}{2} \right) \left( 1 - \frac{\rho_o}{\rho_c} \right) } \\
		\times \left[ 1 + \bar{\mu} - \frac{\rho_o}{\rho_c}
		+ \frac{3}{2} \frac{\rho_o}{\rho_c} \left( 1 - \frac{\rho_o}{\rho_c} \right)
		+ \frac{3}{2} \left( \frac{R_c}{R_p} \right)^5  \left( 1 - \frac{\rho_o}{\rho_c} \right)^2 
		\right] \: .
\end{multline}
Injecting this relation into the expression \eqref{zetaR} of ${\zeta}_c / R_c$, and the resulting relation in the expression \eqref{Fg} of $F$, we finally get:
\begin{equation}
\label{F}
	F = \frac{
		\left( 1 - \frac{\rho_o}{\rho_c} \right) \, (1 + \bar{\mu}) \, \left( 1 + \frac{3}{2 \alpha} \right)
	}{
		1 + \bar{\mu} - \frac{\rho_o}{\rho_c} + \frac{3}{2} \frac{\rho_o}{\rho_c} \left( 1 - \frac{\rho_o}{\rho_c} \right) - \frac{9}{4 \alpha} \left( \frac{R_c}{R_p} \right)^5  \left( 1 - \frac{\rho_o}{\rho_c} \right)^2 
	} \: .
\end{equation}

In the case of a shallow oceanic envelope ($R_p \simeq R_c$), the height of the oceanic tide is then given by $R_c (T_2 - S_2)$ at the surface of the core.
Using Eqs.~(\ref{zetaR}-\ref{S2}), we obtain the classical expression of the height of oceanic tide
\begin{equation}
\label{RTS_shallow}
	R_c (T_2 - S_2) = \frac{ \bar{\mu} {\zeta}_c }
		{ 1- \frac{\rho_o}{\rho_c} + \bar{\mu} \left( 1 - \frac{3}{5} \frac{\rho_o}{\rho_c} \right) } \:.
\end{equation} 

The height of the solid tidal displacement is given by $R_c S_2$.
Using Eqs.~(\ref{RTS_shallow}-\ref{zetaR}-\ref{S2}), we obtain its classical expression:
\begin{equation}
\label{RS_shallow}
	R_c S_2 = \frac{ \frac{5}{2} {\zeta}_c \left( 1 - \frac{\rho_o}{\rho_c} \right) }
		{ 1- \frac{\rho_o}{\rho_c} + \bar{\mu} \left( 1 - \frac{3}{5} \frac{\rho_o}{\rho_c} \right) } \, ,
\end{equation} 
which reduces to
\begin{equation}
	R_c S_2 = \frac{ \frac{5}{2} {\zeta}_p }{1 + \bar{\mu}}
\end{equation}
for an ocean-free planet ($\rho_o = 0$), which corresponds to that given by Lord Kelvin (1863).
Thus, recalling Eq.~\eqref{Fg}, we deduce that for an oceanless planet, $F$ is unity.

\begin{figure}[!htb]
\centering
 \includegraphics[width=\linewidth] {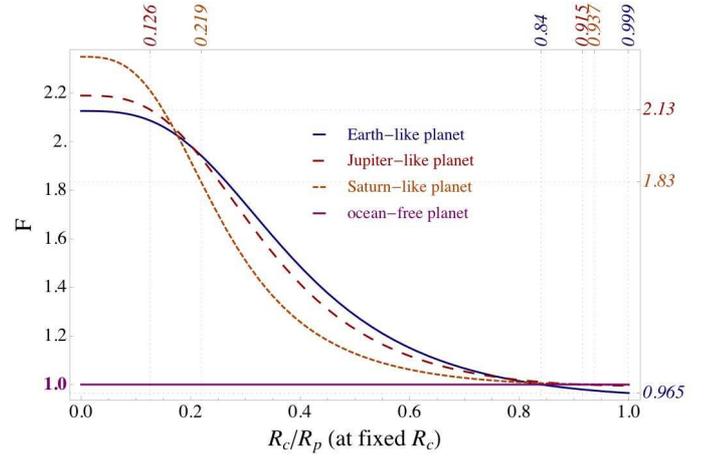}
 \caption{
 Factor $F$, accounting for the overloading exerted by the tidally deformed oceanic envelope on the solid core, in terms of the ocean depth throw the ratio $R_c/R_p$ for three types of planet.
 Parameters are given in Tables~\ref{tab:EarthParams}-\ref{tab:MpRp}-\ref{tab:McRc}: for Earth-, Jupiter- and Saturn-like planets, we assume respectively that: ${R_p = \{1, 10.97 , 9.14\}}$ (in units of $R_p^{\mcal\Phi}$), ${M_p = \{1, 317.8, 95.16\}}$  (in units of $M_p^{\mcal\Phi}$), ${R_c = \{0.99, 0.126 , 0.219\} \times R_p}$, ${M_c = \{0.33, 6.41, 18.65\}}$ (in units of $M_p^{\mcal\Phi}$) and ${\mu = 5 \times 10^{10} \:(\textrm{Pa})}$ for all cores.
 Note that this figure is similar to that drawned by Dermott (1979), divergences coming from the use of different values of the parameters.
 }
 \label{fig:Graph_F}
\end{figure}

Fig.~\ref{fig:Graph_F} displays the value of $F$ for three types of planets ({\it i.e.} Earth-, Jupiter- and Saturn-like planets), with a given core (of fixed size, mass and shear modulus) and a fluid shell of fixed density but variable depth, so that the size and mass of the whole planet varies also. The variation of $F$ is represented in function of $R_c/R_p$: the smaller this ratio, the higher the ocean depth. 

\begin{table}[!Hb]
\begin{tabular}{ c | c | l l }
\hline
\hline
 Quantity & Value & \multicolumn{2}{c}{Reference} \\
\hline
 $M_p^{\mcal\Phi}$ (kg) & $5.9736 \times 10^{24} $             
 	&\hspace*{-0.3cm}\multirow{3}{*}{\begin{minipage}{-0.05\linewidth}$\left. \begin{array}{c} \\ \\ \\ \end{array} \right\}$\end{minipage}} 
 	& \multirow{3}{*}{\begin{minipage}{0.46\linewidth} 
 		{\hspace*{-0.2cm} http://nssdc.gsfc.nasa.gov/ \\ \hspace*{-0.2cm} planetary/factsheet/earthfact.html}
 \end{minipage}} \\
 $R_p^{\mcal\Phi}$ (m)  & $6371 \times 10^3$                   &&  \\
 $M_c^{\mcal\Phi}$ (kg) & $M_p^{\mcal\Phi}-1.4 \times 10^{21}$ &&  \\
 $R_c^{\mcal\Phi}$ (m)  & $R_p^{\mcal\Phi}-3682 \times 10^3$   & \multicolumn{2}{l}{Charette \& Smith (2010)} \\
\hline
\end{tabular}
\caption{\label{tab:EarthParams}Earth parameters.}
\end{table}

\begin{table}[!Hb]
\begin{tabular}{ c | c | c | l }
\hline
\hline
 Quantity& Jupiter & Saturn & \multicolumn{1}{c}{Reference} \\
\hline
 $R_p$ (m) & $10.973 \times R_p^{\mcal\Phi}$ & $9.140 \times R_p^{\mcal\Phi}$  
 & \multirow{2}{*}{\begin{minipage}{0.31\linewidth} 
 		{http://nssdc.gsfc.nasa.\\gov/planetary/factsheet/}
 \end{minipage}}  \\
 $M_p$ (kg) & $317.830 \times M_p^{\mcal\Phi}$  & $95.159 \times M_p^{\mcal\Phi}$ & \\
\hline
\end{tabular}
\caption{\label{tab:MpRp}Mass and mean radius of Jupiter and Saturn.}
\end{table}

For the Earth, all parameters are well known (see Table~\ref{tab:EarthParams}).

\begin{table}[!Htb]
\begin{tabular}{ c | c | c | l }
\hline
\hline
  & $R_c$ (m)  & $M_c$ (kg) & Reference \\
\hline
 \multirow{2}{*}{Jupiter} & \multirow{2}{*}{$0.126 \times R_p$} & \multirow{2}{*}{$6.41 \times M_p^{\mcal\Phi}$} & \multirow{2}{*}{\begin{minipage}{0.2\textwidth}
 	http://www.oca.eu/guillot/ \\jupsat.html (Guillot, 1998)
 \end{minipage}}  \\ &&&\\
 \hline
 Saturn & $0.219 \times R_p$ & $18.65 \times M_p^{\mcal\Phi}$  & Hubbard et al. (2009) \\ 
\hline
\end{tabular}
\caption{\label{tab:McRc}Mass and mean radius of Jupiter's and Saturn's cores.}
\end{table}

The values of the ocean density for the Jupiter- and Saturn-like planets correspond to the ones we may deduce from the well known values of their global size and mass (see Table~\ref{tab:MpRp}), and the much less constrained values of the core size and mass (see models A of Table~\ref{tab:McRc}).
Concerning the shear modulus, we used the value taken as reference by Henning et al. (2009) when studing the tidal heating of terrestrial exoplanets, {\it i.e.} ${\mu = 5 \times 10^{10} \, \mrm{Pa}}$.
These {\it models} of planets are used as starting points to compare the influence the ocean depth has on core deformation for different types of planets.
Since we do not try to estimate this deformation for realistic planets, we will not discuss in this section the validity of the values we use for the parameters.

Fig.~\ref{fig:Graph_F} shows that for a planet with a shallow fluid shell ({\it i.e.} when ${R_c \gtrsim a \, \times \, R_p}$, where ${ a = \{ 0.840 , 0.915 , 0.937 \} }$ for an Earth-, Jupiter- and Saturn-like planet respectively), $F$ is less than unity, which means that the ocean exerts a loading effect on the solid core which is stronger than the gravitational forces and opposed to it.
That is the case of the Earth where the depth of the oceanic envelope does not exceeds $1\%$ of the size of the planet, but giant planets are supposed to have a solid core no bigger than the third of the planet size.
According to Fig.~\ref{fig:Graph_F}, $F$ may reach values up to $2.3$ for this kind of planets.
That means that for a planet with a deep fluid envelope, the ocean tide has no loading effect on the core but exerts a gravitational force that amplifies the tidal deformation.
We refer the reader to Dermott (1979) for a complete discussion.

\subsection{Modified Love numbers}

From Eq.~\eqref{F} we deduce the Love number $h^F_2$ (cf. Eq.~\ref{h2'}) which measures the surface deformation:
\begin{equation}
	h^F_2 = \frac{ \frac{5}{2} \, 
		\left( 1 - \frac{\rho_o}{\rho_c} \right) \, \left( 1 + \frac{3}{2 \alpha} \right)
	}{
		1 + \bar{\mu} - \frac{\rho_o}{\rho_c} + \frac{3}{2} \frac{\rho_o}{\rho_c} \left( 1 - \frac{\rho_o}{\rho_c} \right) - \frac{9}{4 \alpha} \left( \frac{R_c}{R_p} \right)^5  \left( 1 - \frac{\rho_o}{\rho_c} \right)^2 
	} \: .
\end{equation}

Let us give here the expression of the second-order Love number \eqref{k2def} associated with the solid core.
First of all, let us recall its value for an ocean-free planet:
According to Eq.~\eqref{Phic1} with $r=R_c$, we get
\begin{equation}
\label{k2}
	k_2 = \frac{3}{5} \, h_2 = \frac{3}{2} \, \frac{1}{1+\bar{\mu}} \: .
\end{equation} 
In presence of an ocean on top of the solid core, we may also introduce the modified Love number
\begin{equation}
\label{k2'_def}
	k^F_2  \stackrel{\text{def}}{=} \frac{\Phi'(R_c)}{U(R_c)} = \frac{V'(R_c)-U(R_c)}{U(R_c)} = \frac{V'(R_c)}{U(R_c)} - 1 \: ,
\end{equation}
where $V'(R_c)$ and $U(R_c)$ are obtained from Eqs.~\eqref{Vc'} and \eqref{U2bis} respectively, with $r=R_c$.
Thus, expressing ${\zeta_c}/{R_c}$ in function of the modified second-order Love number $h_2^F$ according to Eq.~\eqref{h2'}, and using Eq.~\eqref{S2}, we obtain the expression of $k^F_2$ in terms of $h_2^F$:
\begin{equation}
\label{k2'}
	k^F_2 = \left( 1 + \frac{2}{5} \, \frac{ \bar{\mu} }{ 1-\frac{\rho_o}{\rho_c} } \right) \, h^F_2 - 1 \:.
\end{equation}

In this section we have studied the impact of the presence of a fluid envelope in the determination of the deformation imposed on an elastic core under tidal forcing.
In the following, we consider that the solid core also presents viscous properties so that its response to the tidal force exerced by the perturber is no more immediate, thus inducing dissipation.
The next section addresses the quantification of this conversion of energy, which drives the dynamical evolution of the whole system.


\section{Anelastic tidal dissipation: analytical results}

Assuming that the anelasticity is linear, the correspondence principle established by Biot (1954) allows us to extend the formulation of the adiabatic elastic problem to the resolution of the equivalent dissipative anelastic problem.
For initial conditions taken as zero and similar geometries, the Laplace and Fourier transforms of the anelastic equations and boundary conditions are identical to the elastic equations, if the rheological parameters and radial functions are defined as complex numbers.
We will then denote 
\begin{equation}
	\tenseurcomplex{\sigma} \equiv \tenseur{\sigma}_1 + \pmb i \, \tenseur{\sigma}_2
\end{equation}
the complex stress tensor, and
\begin{equation}
	\tenseurcomplex{\epsilon} \equiv \tenseur{\epsilon}_1 + \pmb i \, \tenseur{\epsilon}_2
\end{equation}
the complex strain tensor.

The perturbative strain is cyclic, with tidal pulsations $\sigma_{2,m,p,q}$.
For sake of clarity, we will use from now on the generic notation ${\omega \equiv \sigma_{2,m,p,q}}$, recalling that there is a large range of tidal frequencies for each term of the expansion of the tidal potential.
The stress and strain tensors take the following form:
\begin{subequations}
\begin{align}
	\tenseurcomplex{\sigma}(\omega) &= (\tenseur{\sigma}_1 + \pmb i \tenseur{\sigma}_2) \, \pmb e ^{\pmb i \omega t} \: ,\\
	\tenseurcomplex{\epsilon}(\omega) &= \tenseurcomplex{\epsilon}_0 \, \pmb e ^{\pmb i \omega t} \: .
\end{align}
\end{subequations}

The complex rigidity
\begin{equation}
\label{mu_tilde}
	\tilde{\mu}(\omega) \equiv {\mu}_1(\omega) + \pmb i \, {\mu}_2(\omega) \:,
\end{equation}
where ${\mu}_1$ represents the energy storage and ${\mu}_2$ the energy loss of the system, is defined by
\begin{equation}
\label{mu_tilde_def}
	\tilde{\mu}(\omega) \equiv \frac{\tenseurcomplex{\sigma}(\omega)}{\tenseurcomplex{\epsilon}(\omega)} \: .
\end{equation}
We may also define the complex effective rigidity
\begin{equation}
\label{mu_bar_tilde}
	\hat{\mu}(\omega) \equiv \bar{\mu}_1(\omega) + \pmb i \, \bar{\mu}_2(\omega)
\end{equation}
by:
\begin{equation}
\label{mu_bar_tilde_def}
	\hat{\mu}(\omega) = \gamma \, \tilde{\mu}(\omega) \:,
\end{equation}
where (see Eq.~\ref{mu_bar})
\begin{equation}
	\gamma = \frac{\hat{\mu}}{\tilde{\mu}} \equiv \frac{\bar{\mu}}{\mu} = \frac{19}{2 \rho_c g_c R_c} \:.
\end{equation}

\subsection{Case of a fully-solid planet}

The complex Love number $\tilde{k}_2$  may be expressed in terms of the complex effective rigidity $\hat{\mu}$, by:
\begin{equation}
\label{k2i}
\begin{split}
	\tilde{k}_2(\omega)
		&= \frac{3}{2} \, \frac{1}{1+ \hat{\mu}(\omega)} \\
		&= \frac{3}{2} \, \frac{1}{1+ \gamma \left[ \mu_1(\omega) + \pmb i \mu_2(\omega) \right] } \: ,
\end{split}
\end{equation}
in the case of a completely solid planet.

The real part of $\tilde{k}_2$ characterizes the purely elastic deformation, since $- \Im(\tilde{k}_2)$ gives the phase lag due to the viscosity.
Therefore, we define the factor of tidal dissipation $Q$, that represents the dissipation rate due to viscous friction, by
\begin{equation}
\label{Q_def}
	Q^{-1}(\omega) = - \frac{\Im{\tilde{k}_2}(\omega)}{\left| \tilde{k}_2(\omega) \right|} \: .
\end{equation}
Then, from equation \eqref{k2i}, we deduce that
\begin{equation}
\label{Q}
	Q(\omega) = \sqrt{ 1 + \left[ \frac{1}{\bar{\mu}_2(\omega)} + \frac{\bar{\mu}_1(\omega)}{\bar{\mu}_2(\omega)} \right]^2 }  \: .
\end{equation}

\subsection{Case of a two-layer planet}

Let us introduce the quantities
\begin{subequations}
\label{AB}
 \begin{flalign}
    A &= \left( 1 - \frac{\rho_o}{\rho_c} \right) \, \left( 1 + \frac{3}{2 \alpha} \right) \:,\:\text{ and: } \label{A}\\
    B &= 1 - \frac{\rho_o}{\rho_c} + \frac{3}{2} \frac{\rho_o}{\rho_c} \left( 1 - \frac{\rho_o}{\rho_c} \right) - \frac{9}{4 \alpha} \left( \frac{R_c}{R_p} \right)^5  \left( 1 - \frac{\rho_o}{\rho_c} \right)^2  \:, \label{B}
 \end{flalign}
\end{subequations}
in the expression \eqref{F} of F:
\begin{equation}
	F = \frac{ A (1+\bar{\mu}) }{B+\bar{\mu}} \: ,
\end{equation}
and let us define its complex equivalent
\begin{equation}
\label{F'}
	\widetilde{F} = \frac{ A (1+\hat{\mu}) }{B+\hat{\mu}} \:.
\end{equation}

The determination of how the presence of an oceanic envelope will modify the tidal dissipation consists in the determination of $\tilde{k}^F_2$, defined as the complex Love number $\tilde{k}_2$ in presence of the fluid envelope.
According to the correspondence principle, this number is given by the complex Fourier transform of \eqref{k2'}, i.e. :
\begin{equation}
	\tilde{k}^F_2(\omega) = \left( 1 + \frac{2}{5} \, \frac{ \hat{\mu}(\omega) }{ 1-\frac{\rho_o}{\rho_c} } \right) \, \tilde{h}^F_2(\omega) - 1 \:.
\end{equation}
From \eqref{k2'} we get then
\begin{multline}
\label{k2i'}
	\tilde{k}^F_2(\omega) = \frac{1}{ \left(B+\bar{\mu}_1 \right)^2 + \bar{\mu}_2^2 } \\
		\times \left\lbrace  \left[  \left(B+\bar{\mu}_1 \right) \, \left( C+\frac{3}{2\alpha} \, \bar{\mu}_1 \right) + \frac{3}{2\alpha} \, \bar{\mu}_2^2 \right] 
	     - \pmb i {A \, D \, \bar{\mu}_2}  \right\rbrace \:,
\end{multline} 
where we made use of the dimensionless quantities $\alpha$, $A$ and $B$ previously defined (see respectively Eqs.~\ref{alpha} and \ref{AB}), $C$ and $D$ given by:
\begin{subequations}
\begin{flalign}
	C &= \frac{3}{2} \, \left( 1 - \frac{\rho_o}{\rho_c} \right) \, \left( 1 - \frac{\rho_o}{\rho_c} + \frac{5}{2\alpha} \right) 
	     + \frac{9}{4\alpha} \left(\frac{R_c}{R_p}\right)^5 \, \left( 1 - \frac{\rho_o}{\rho_c} \right)^2 \:, \\
	D &= \frac{3}{2} \, \left( 1 - \frac{\rho_o}{\rho_c} \right) \, \left[ 1 + \frac{3}{2\alpha} \, \left(\frac{R_c}{R_p}\right)^5 \right] \:.
\end{flalign}
\end{subequations}
Finally, the dissipation factor $\hat{Q}$, defined here by
\begin{equation}
\label{Q'_def}
	\hat{Q}^{-1}(\omega) = - \frac{\Im{\tilde{k}^F_2(\omega)}}{ \left| \tilde{k}^F_2(\omega) \right|} \: ,
\end{equation} 
is of the form:
\begin{equation}
\label{Q'}
	\hat{Q}(\omega) = 
	\sqrt{ 1 + \frac{9}{4 \alpha^2 A^2 D^2} \, 
	       \left[ 1 + \frac{\left( B + \bar{\mu}_1(\omega) \right) \, \left( \frac{2 \alpha C}{3} + \bar{\mu}_1(\omega) \right) }{ \bar{\mu}_2(\omega) } \right]^2 } \: .
\end{equation}

Thanks to the correspondence principle, one is able to derive this general expression of the tidal dissipation valid for any rheology.
The obtained formulae explicitly reveal its dependence on the tidal frequency ${\omega \equiv \sigma_{2,m,p,q}}$, as shown, for example, by Remus et al. (2012), Ogilvie \& Lin (2004-2007) for fluid layers.

\subsection{Implementation of an anelastic model}

Since the factor $Q$ depends on the real and imaginary components $\bar{\mu}_1$ and $\bar{\mu}_2$ of the complex effective shear modulus $\hat{\mu}$, we need to define the rheology of the studied body to express it in terms of the constitutive parameters of the material.

The {\it anelasticity} of a material is evaluated by a quality factor $Q_a$ defined by
\begin{equation}
	\label{Qa_def}
	Q_a(\omega) = \frac{\mu_1(\omega)}{\mu_2(\omega)} \: .
\end{equation}
We may express the solid tidal dissipation, given by $Q$ (Eq.~\ref{Q}) for a fully-solid planet and $\hat{Q}$  (Eq.~\ref{Q'}) in the case of a two-layer planet, in terms of this factor:
\begin{equation}
\label{Q_model}
	Q(\omega \equiv \sigma_{2,m,p,q}) = \sqrt{ 1 + \left[ \frac{1}{\bar{\mu}_2(\omega)} + Q_a(\omega) \right]^2 }  \:,
\end{equation}
and
\begin{multline}
\label{Q'_model}
	\hat{Q}(\omega \equiv \sigma_{2,m,p,q}) = \\
	\sqrt{ 1 + \frac{9}{4 \alpha^2 A^2 D^2} 
	   \left[ 1 + \left( \frac{B}{\bar{\mu}_2(\omega)} + Q_a(\omega) \right) 
	   	   \, \left( \frac{2}{3} \frac{ \alpha C}{\bar{\mu}_2(\omega)} + Q_a(\omega) \right) \right]^2 } \: .
\end{multline}

All previous results are independent of the viscoelastic rheological model.
We have now to apply these general expressions to a specific rheology which will depend on the physical properties of the material.
Considering our lack of knowledge on the internal structure of giant planets, we will implement the simplest model, namely the Maxwell model, 
which presents the advantage to involve only two parameters and thus to be easy to use (Tobie 2003, and Tobie et al. 2005).
A critical overview of the four main rheological models has also been done by Henning et al. (2009), thus we refer the reader to the three above mentioned papers for a detailed comparison.

\subsection{The Maxwell model.\\}

\begin{figure}[!htb]
      \centering
      \includegraphics[width=0.8\linewidth]{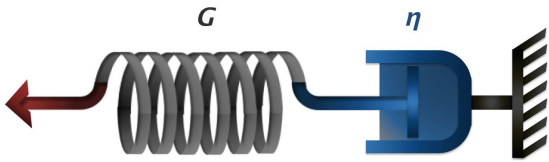}
      \caption{
      Representation of the Maxwell model and its corresponding notations.
      \label{fig:Maxwell}
      }
\end{figure}

This model considers a viscoelastic material as a spring-dashpot serie.
The instantaneous elastic response is characterized by a shear modulus $G$ , and the viscous yielding is represented by a viscous scalar modulus $\eta$ (see Fig.~\ref{fig:Maxwell}).
Notice that the shear moduli $G$ and $\mu$ (introduced in Sect.~\ref{subsection:dyn_eq_lin}) designate the same quantity.
We change here the notation to avoid any confusion with the complex shear modulus $\tilde{\mu}$ used to study the anelastic tidal dissipation, and whose real and imaginary parts involve both $G$ and $\eta$.

\begin{figure}[!htb]
\centering
 \includegraphics[width=\linewidth] {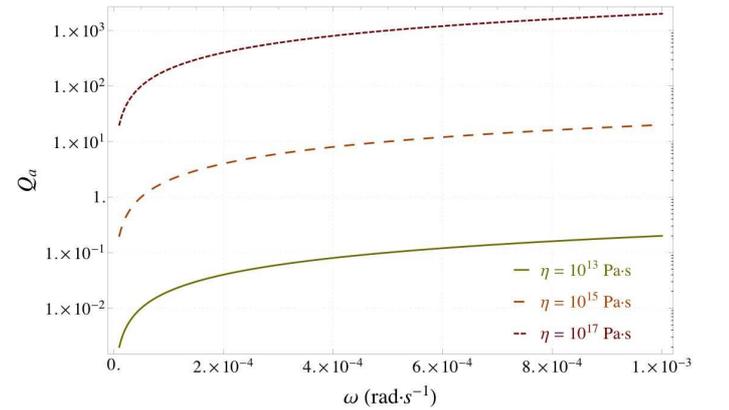}
 \caption{
 Anelastic quality factor $Q_a$ of the Maxwell model in function of the tidal pulsation $\omega$ for different values of the viscosity $\eta$. G is taken equal to ${5 \times 10^{10} ~\mrm{Pa}}$ (see Henning et. al 2009). $Q_a$ is represented on a logarithmic scale.
 }
 \label{fig:Graph_Qa}
\end{figure}

The constitutive equation is given by Henning et al. (2009):
\begin{equation}
\label{eq_const}
	G \, \tenseurcomplex{\sigma}(\omega) + \eta \, \dot{\tenseurcomplex{\sigma}}(\omega) 
	= G \, \eta \, \dot{\tenseurcomplex{\epsilon}}(\omega) \: ,
\end{equation}
where the time derivative of a given quantity is denoted by a dot.
Recalling the relation \eqref{mu_tilde_def} this equation becomes:
\begin{equation}
	G \, \tenseurcomplex{\epsilon}(\omega) \, \tilde{\mu}(\omega) 
	+ \eta \, \frac{\mrm d}{\mrm dt} \left[ \tilde{\mu}(\omega) \, \tenseurcomplex{\epsilon}(\omega) \right] 
	= G \, \eta \, \dot{\tenseurcomplex{\epsilon}}(\omega) \: .
\end{equation}
Therefore the real part $\mu_1$ and the imaginary part $\mu_2$ of the complex shear modulus $\tilde{\mu}$ have the following expressions:
\begin{subequations}
\label{mu12}
\begin{flalign}
	& \mu_1(\omega) = \frac{{\eta}^2 \, G \, \omega^2}{G^2 + {\eta}^2 \, \omega^2} \:, \label{mu1} \\
	& \mu_2(\omega) = \frac{\eta \, G^2 \, \omega}{G^2 + {\eta}^2 \, \omega^2} \: . \label{mu2}
\end{flalign}
\end{subequations}
The {\it anelastic quality factor} $Q_a$ is then given by
\begin{equation}
\label{Qa}
	Q_a (\omega) = \frac{\mu_1(\omega)}{\mu_2(\omega)} = \frac{\eta \, \omega}{G} \equiv \omega \, \tau_M \:,
\end{equation}
where $\tau_M = \eta / G $ is the characteristic time of relaxation of the Maxwell model.
As confirmed by Fig.~\ref{fig:Graph_Qa}, Eq.~\eqref{Qa} shows that $Q_a$ increases linearly with the frequency of the cyclic tidal strain: the shorter the oscillation period, the lower the dissipation due to intrinsic viscoelastic properties of the material.
Moreover, the anelastic quality factor is proportional to $\tau_M = \eta / G$, so that it dissipates more if it is more rigid and less viscous.

Thus, we may express $\mu_2$ (Eq.~\ref{mu2}) in terms of the anelastic quality factor $Q_a$ (Eq.~\ref{Qa}):
\begin{equation}
\label{mu2_Qa}
	\mu_2(\omega) = \frac{G}{\left( \omega \, \tau_M \right)^{-1} + \omega \, \tau_M} \:.
\end{equation}

In the case of a fully-solid body, we get, from Eqs.~\eqref{Qa_def} and \eqref{mu2_Qa}, the imaginary part of the complex Love number $\tilde{k}_2$ \eqref{k2i}:
\begin{equation}
\label{Imk2_max}
	\Im \left[ \tilde{k}_2(\omega) \right] 
	= - \frac{3 \gamma}{2} \, \frac{G \, \omega \, \tau_M}{1 + \left({\omega \, \tau_M} \right)^2 \, ( 1 + \gamma G )^2 } \:.
\end{equation}
Therefore the dissipation factor $Q$ defined by \eqref{Q} is of the form:
\begin{multline}
\label{Q_max}
	Q(\omega \equiv \sigma_{2,m,p,q}) = \\
	\sqrt{ 1 + \left\{ \frac{1}{G \gamma} \, \left[ \left( \omega \, \tau_M \right)^{-1} + \omega \, \tau_M \right] + \omega \, \tau_M \right\}^2 } \:.
\end{multline}

In the more general case of a two-layer body, the imaginary part of the complex Love number $\tilde{k}^F_2$ given by Eq.~\eqref{k2i'}, takes a different form than Eq.~\eqref{Imk2_max} because of the presence of the fluid envelope, so does the two-layer dissipation factor $\hat{Q}$ (\ref{Q'}) with respect to its oceanless form \eqref{Q_max}.
To obtain them, one needs to replace the shear modulus $\tilde{\mu}$ and the anelastic quality factor $Q_a$ by their expression in the case of the Maxwell model (Eqs~\ref{mu12} and \ref{Qa} respectively).

\begin{figure}[!htb]
\centering
 \includegraphics[width=\linewidth] {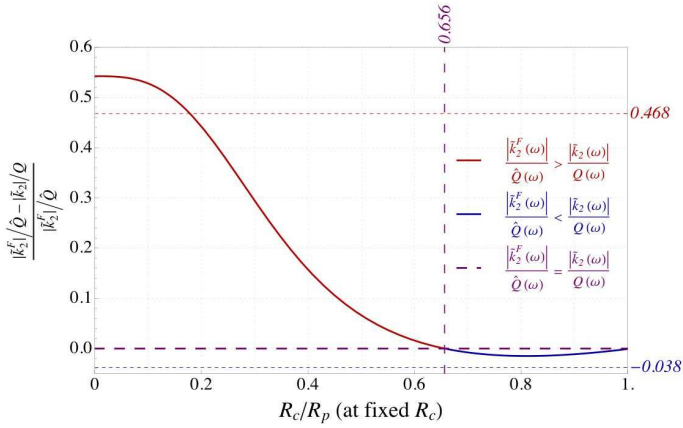}
 \caption{
 Relative difference between ${{\left|\tilde{k}^F_2\right|}/{\hat{Q}}}$ and ${{\left|\tilde{k}_2\right|}/{Q}}$. 
 We may distinguish two regimes: for $0<R_c<0.661 \, R_p$ the presence of the fluid envelope increases the tidal dissipation; over this value, tidal dissipation is lower than it would have been without fluid shell.
Parameters are given in Tables~\ref{tab:MpRp}-\ref{tab:McRc}-\ref{tab:omega} for a Saturn-like planet perturbed at the tidal frequency of Enceladus (${\omega = 2.25 \times 10^{-4} \, \mrm{rad}\cdot \mrm{s}^{-1}}$): ${R_p = 9.14 \, R_p^{\mcal\Phi}}$, ${M_p = 95.16 \, M_p^{\mcal\Phi}}$, ${R_c = 0.219 \, R_p^{\mcal\Phi}}$, ${M_c = 18.65 \, M_p^{\mcal\Phi}}$, ${G = 5 \times 10^{10} \:(\textrm{Pa})}$ and ${\eta = 10^{15}} \, (\mrm{Pa}\cdot\mrm{s})$.
 }
 \label{fig:Graph_k2Q_comp}
\end{figure}

Fig.~\ref{fig:Graph_k2Q_comp} compares the dissipation of the solid core with and without the presence of a fluid envelope of variable depth for a Saturn-like planet, using the parameters given by Tables~\ref{tab:MpRp}-\ref{tab:McRc}-\ref{tab:omega}: as expected, the difference between the two decreases with the size of the fluid envelope down to about ${0.34 \times R_p}$; but for a thiner fluid shell, the dissipation get lower than it would be without it. 

\begin{table}[!h]
\begin{tabular}{ c | c | c }
\hline
\hline
 Planet &Jupiter & Saturn \\
\hline
Satellite & Io & Enceladus \\
\hline
 $\omega \text{ (rad} \cdot \text{s}^{-1})$ &$2.79 \times 10^{-4}$ & $2.25 \times 10^{-4}$ \\
 \hline
Reference & Ioannou et Lindzen (1993) & Lainey et al. (2012)  \\
\hline
\end{tabular}
\caption{\label{tab:omega}Tidal frequencies considered in numerical applications.}
\end{table}


\section{Anelastic tidal dissipation: role of the structural and rheological parameters}

With our choice of the Maxwell model to represent the rheology of the solid parts of the planet, the dissipation quality factor $\hat{Q}$ depends on the tidal frequency $\omega$ and on four structural and rheological parameters: the relative size of the core ($R_c / R_p$), the relative density of the envelope with respect to the core ($\rho_o / \rho_c$), the shear modulus ($G$) and the viscosity of the core ($\eta$).
Our present knowledge of extrasolar giant planets, and also planets of our Solar System like Jupiter or Saturn, suffers some uncertainties on the values of these parameters, so that the range of values taken by core properties of giant planets presents poor constraints.
Moreover, even if the presence of a core in Jupiter is not yet confirmed (see Guillot 1999-2005), but new data coming from seismology may provide more constraints on giant planets internal structure (see Gaulme et al. 2011).
Nevertheless, we will explore the resulting tidal dissipation of such bodies around values of the structural and rheological parameters taken as reference and corresponding to those of the literature.

\subsection{Baseline structural and rheological parameters}
\label{sect:params}

As reference models, we chose Jupiter and Saturn although their core parameters are still uncertain.
The values of the global sizes and masses of these planets are those of Table~\ref{tab:MpRp}.

\subsubsection{Size and mass of the core}
\label{Sect:core_struct_params}
Presently, two main types of models are available for Jupiter's interior.
The NHKFRB group\footnote{Nettelmann, Holst, Kietzmann, French, Redmer and Blaschke (Nettelmann et al. 2008).} uses a three-layer model with a thin radiative zone, close to previous models by Saumon \& Guillot (2004), whereas the MHVTB group\footnote{Militzer, Hubbard, Vorberger, Tamblyn, and Bonev (Militzer et al. 2008).} proposes a new type of Jupiter model that possesses only two layers (see Militzer et al. 2008).
But, as explained in Militzer \& Hubbard (2009), the crucial difference lies in the treatment of the molecular-to-metallic transition in dense fluid hydrogen, that leads to very different conclusions.
The first group predicts a small core of less than 10 $M_p^\mcal{\Phi}$ (Saumon \& Guillot 2004), while the second obtains a larger core of 14-18 $M_p^\mcal{\Phi}$ (Militzer et al. 2008).
Among all these models of Jupiter's interior, we choose as reference the adiabatic model with Plasma Phase Transition (PPT) (http://www.oca.eu/guillot/jupsat.html)\footnote{It is constructed with CEPAM, Code d'Evolution Plan\'etaire Adaptatif et Modulaire (Guillot \& Morel, 1995).} of Guillot (1998), which is of the first type.
It predicts a core of radius  ${R_c = 0.126 \times R_p}$ and mass ${M_c = 6.41 \times M_p^\mcal{\Phi}}$.
Only the mass of the core of this reference model will be used in what follows.
The core radius will serve just as a first approximation, as a starting point in our study, since we will present our results for several core sizes.

\begin{figure*}[!htb]
\centering
 \includegraphics[width=\linewidth] {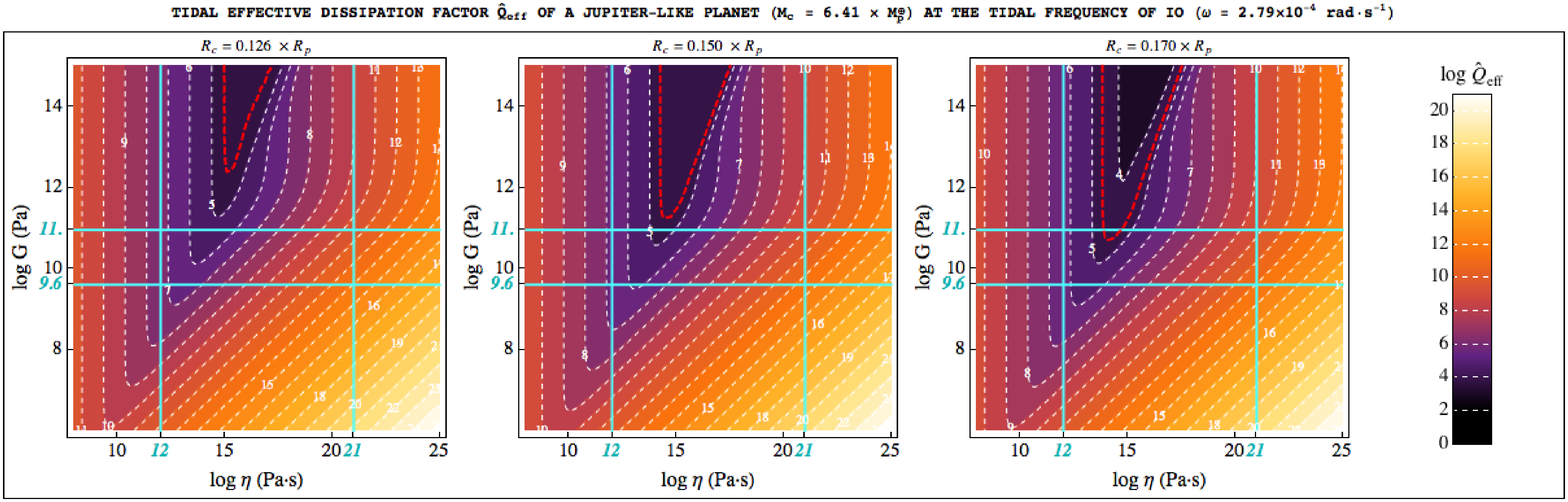}
 \includegraphics[width=\linewidth] {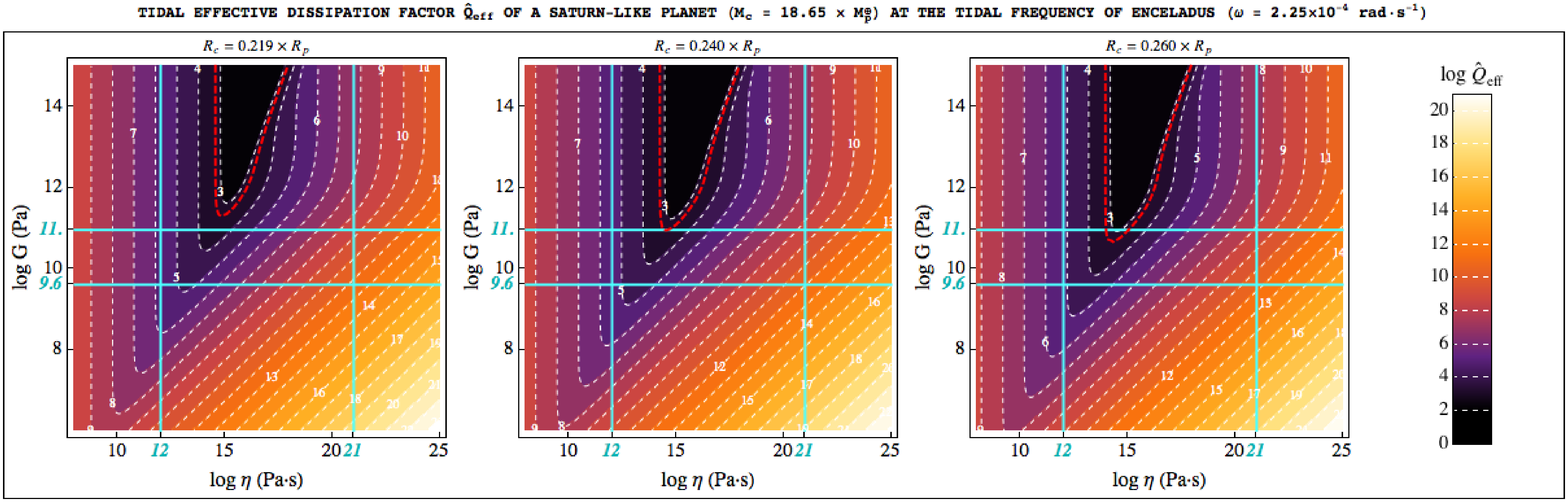}
 \caption{
  Dissipation quality factor $\hat{Q}_\mrm{eff}$ of the Maxwell model in function of the viscoelastic parameters $G$ and $\eta$.
  {\it Top}: for a Jupiter-like two-layer planet tidally perturbed at the Io's frequency ${\omega \simeq 2.79 \times 10^{-4} ~\mrm{rad \cdot s^{-1}}}$.
  {\it Bottom}: for a Saturn-like two-layer planet tidally perturbed at the Enceladus' frequency ${\omega \simeq 2.25 \times 10^{-4} ~\mrm{rad \cdot s^{-1}}}$.
  The red dashed line corresponds to the value of $\hat{Q}_\mrm{eff}=\{(3.56 \pm 0.56) \times 10^4,(1.682 \pm 0.540) \times 10^3\}$ (for Jupiter and Saturn respectively) determined by Lainey et al. (2009-2012).
  The blue lines corresponds to the lower and upper limits of the reference values taken by the viscoelastic parameters $G$ and $\eta$ for an unknown mix of ice and silicates.
  We assume the values of ${R_p = \{10.97 , 9.14\}}$ (in units of $R_p^{\mcal\Phi}$), ${M_p = \{317.8, 95.16\}}$  (in units of $M_p^{\mcal\Phi}$), ${M_c = \{6.41, 18.65\} \times M_p^{\mcal\Phi}}$ given in Tables~\ref{tab:MpRp}-\ref{tab:McRc}.
 }
 \label{fig:Graph_Qeff_rheology}
\end{figure*}

There are also different models of Saturn's interior.
According to the model of Guillot with PPT (http://www.oca.eu/guillot/jupsat.html)\footnotemark[3], Saturn's core may have a mass of ${M_c = 6.55 \times M_p^\mcal{\Phi}}$ and a size of ${R_c = 0.174 \times R_p}$.
More recently, Hubbard et al. (2009) infer, from Cassini-Huygens data, that Saturn has a larger core in the range $M_c = \text{15-20} \times M_p^{\mcal\Phi}$ and a corresponding radius of more than $20\%$ of the planet size. 
We adopt this latter as reference model of Saturn, with ${M_c = 18.65 \times M_p^{\mcal\Phi}}$ and ${R_c = 0.219 \times R_p}$.

All these models support core accretion as the standard process for the formation of giant planets.
The corresponding parameters are listed in Table~\ref{tab:McRc}.

\subsubsection{Rheological parameters of the core}
\label{Sect:core_rheolog_params}
The main uncertainties concern the viscoelastic properties of the core, namely its shear modulus $G$ and its viscosity $\eta$.
At high pressure and temperature, theoretical models and experiments show that $G$ and $\eta$ values depend on temperature and pressure.
But no experiments are available at the very-high pressures and temperatures we may expect in Jupiter's and Saturn's cores (Guillot 2005).
Nevertheless, geophysical and experimental data have allowed to constrain the rheology of the icy satellites of Jupiter, since their ranges of pressure and temperature are similar to those of the outer mantle of the Earth (Tobie 2003).
Then, keeping in mind that these values may differ by several orders of magnitude in our case, we will adopt reference values based on these data, assuming that Jupiter's and Saturn's cores are made of ice and rock.
We will then explore, in the following Sect.~\ref{sect:rheology}, the variation of the tidal dissipation for a large range of values of the rheological parameters around those taken as reference.

We thus assume that the shear modulus $G$ is in the range ${ \left[ G_{ice} = 4\times10^9 \,(\mrm{Pa}) \,,\, G_{silicate} = 10^{11} \,(\mrm{Pa}) \right]}$ (Henning et al. 2009).

Concerning the viscosity $\eta$, it takes its values in the range ${\left[\eta_{ice}=10^{14}\mrm{Pa}\cdot\mrm{s},\eta_{silicate}=10^{21}\,\mrm{Pa}\cdot\mrm{s}\right]}$ for the icy satellites of Jupiter at high pressure (Tobie 2003).
We expand this range, reducing its lower boundary by two orders of magnitude, following the discussion of Karato (2011) which seems to indicate that, at the very high pressures, viscosity in deep interior of super-Earths may decrease by two or three orders of magnitude.
We refer the reader to the Karato's paper for an overview of all plausible mechanisms that may cause a change in the viscous-pressure relationship at very-high pressures.

\begin{figure*}[!htb]
\centering
 \includegraphics[width=0.49\linewidth] {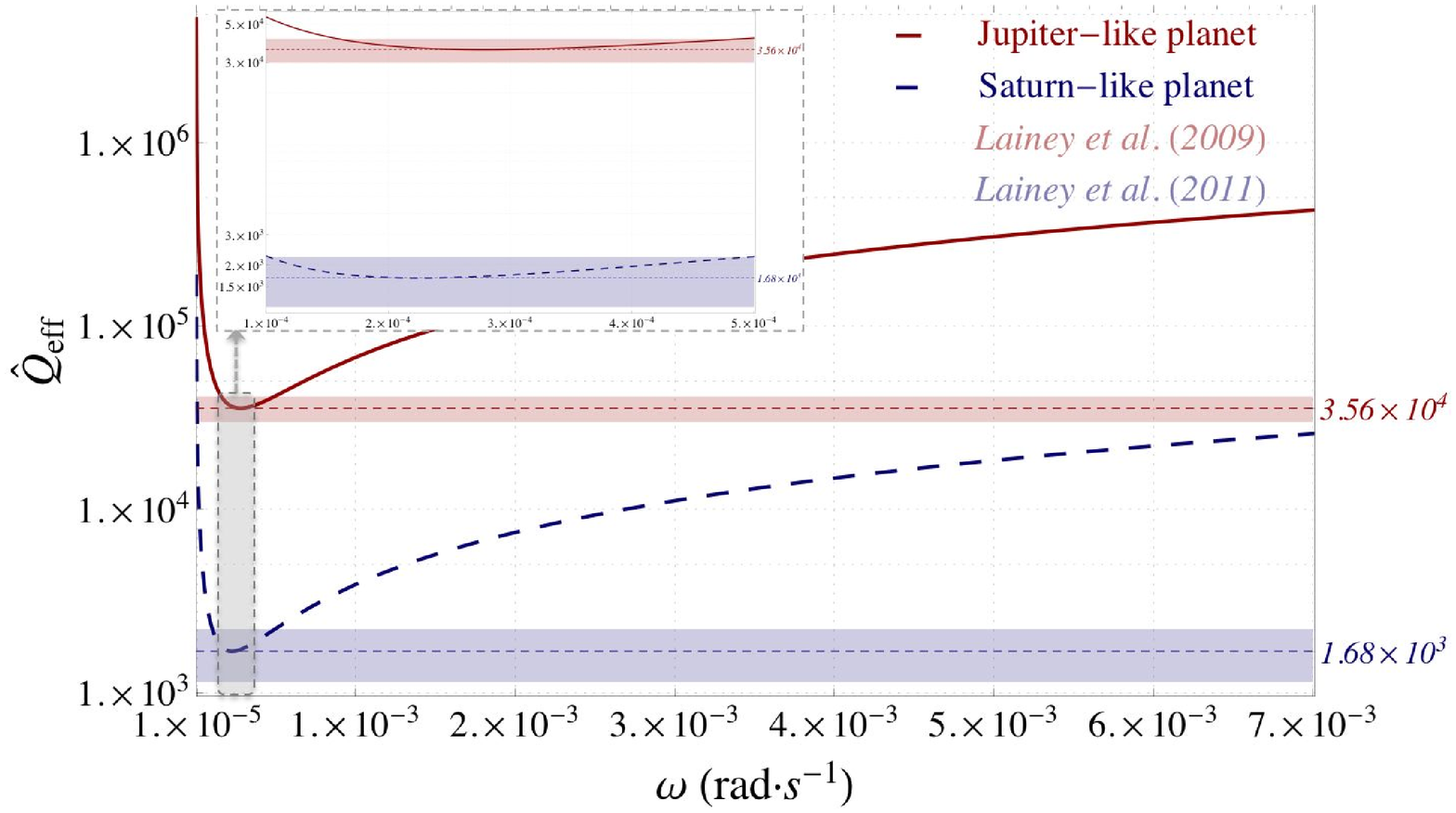}
 \includegraphics[width=0.49\linewidth] {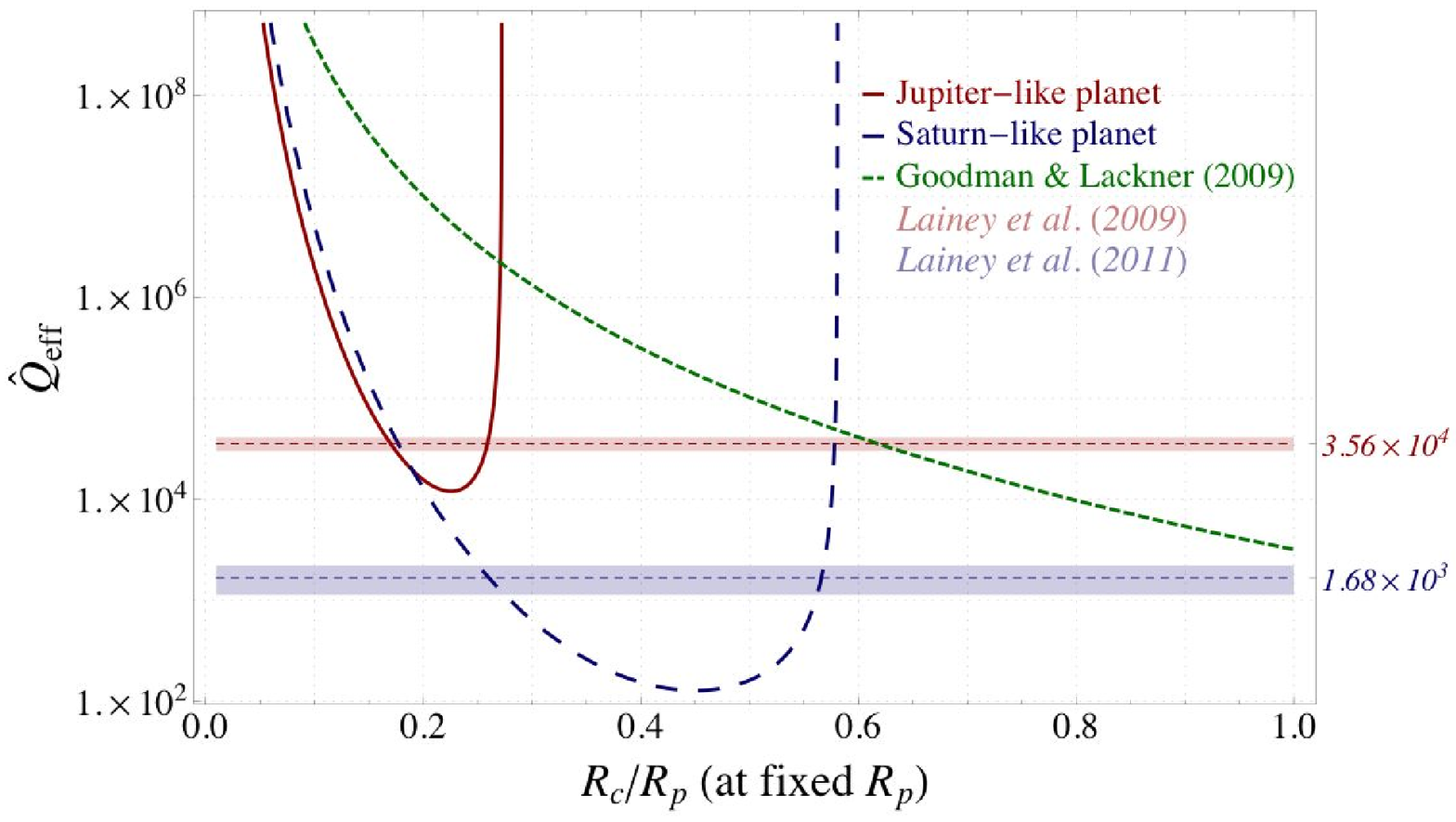}
 \caption{\label{fig:Graph_Q1}
	  Dissipation quality factor ${\hat{Q}_\mrm{eff}}$ normalized to the size of the planet for Jupiter-like and Saturn-like giant planets.
	  Note that all curves are represented with a logarithmic scale.
	  {\it Left}: dependence to the perturbative strain pulsation $\omega$, with ${R_c = \{0.20 , 0.34\} \times R_p}$.
	  {\it Right}: dependence to the size of the core, with ${\omega \simeq 2.25 \times 10^{-4} ~\mrm{rad \cdot s^{-1}}}$ (tidal frequency of Enceladus) for the blue curve associated to a Saturn-like planet, and ${\omega \simeq 2.79 \times 10^{-4} ~\mrm{rad \cdot s^{-1}}}$ (tidal frequency of Io) for the red curve associated to a Jupiter-like planet.
	  The green curve corresponds to the prescription of Goodman \& Lackner (2009) (see plain text for details).
  The red and blue dashed lines correspond to the value of $\hat{Q}_\mrm{eff}=\{(3.56 \pm 0.56) \times 10^4,(1.682 \pm 0.540) \times 10^3\}$ (for Jupiter and Saturn respectively) determined by Lainey et al. (2009-2012).
 Their zone of uncertainty is also represented in the corresponding color.
  We assume the values of ${R_p = \{10.97 , 9.14\}}$ (in units of $R_p^{\mcal\Phi}$), ${M_p = \{317.8, 95.16\}}$  (in units of $M_p^{\mcal\Phi}$), ${M_c = \{6.41, 18.65\} \times M_p^{\mcal\Phi}}$ given in Tables~\ref{tab:MpRp}-\ref{tab:McRc}.
  We also take for the viscoelastic parameters ${G= \{4.85 , 4.45\} \times 10^{10} \, \textrm{(Pa)}}$, and ${\eta = \{1.26 , 1.78\} \times 10^{14} \, (\textrm{Pa}\cdot{\mrm{s}^{-1}})}$ for Jupiter and Saturn respectively. }
\end{figure*}
\subsection{Dependence of tidal dissipation on rheology}
\label{sect:rheology}

Since tidal dissipation causes exchange of angular momentum in the system, it may be quantified by monitoring carefuly the orbital motion of the system.
Using astrometric data covering more than a century, Lainey et al. (2009-2012) succeeded in determining from observations the tidal dissipation in Jupiter and Saturn: namely, ${Q_\mrm{Jupiter} = (3.56 \pm 0.56) \times 10^4}$ determined by Lainey et al. (2009), and ${Q_\mrm{Saturn} = (1.682 \pm 0.540) \times 10^3}$ determined by Lainey et al. (2012) and requireded by the formation scenario of Charnoz et al. (2011).
However, with such a method, the different contributions to the global tidal dissipation, coming from each layer constituting the planet, are lumped together.
Equations of the dynamical evolution (Eqs.~\ref{Evol_Omega} to \ref{Evol_i}) link the observed evolution rates of the rotational and orbital parameters to the observed tidal dissipation and system characteristics.
Since all these rates are proportional to $R_p^5$, where $R_p$ is the planet radius, we introduce the associated dissipation factor \footnote{This can also be demonstrated by calculating the perturbation of the gravific potential at the surface of the planet, adapting Zahn (1966) theory taking into account the boundary conditions at the core surface.}:
\begin{equation}
	\hat{Q}_\mrm{eff} = \left( \frac{R_p}{R_c} \right)^5 \times \frac{|\tilde{k}_2^F(R_p)|}{|\tilde{k}_2^F(R_c)|} \times \hat{Q}  \:,
\end{equation}
where $|\tilde{k}_2^F(R_c)|$ can be deduced from Eq. \eqref{Q'_def}, and $|\tilde{k}_2^F(R_p)|$ designates the modulus of the second order Love number of the planet's surface that is obtained from Eqs. \eqref{Pot_tot_sp} and \eqref{U2bis}
\begin{equation}
\begin{split}
	\tilde{k}_2^F(R_p) &= \frac{V'(R_p)}{U(R_p)} - 1  \\
				 &=  \frac{ \frac{\rho_o}{\rho_c} - \left( \frac{R_c}{R_p} \right)^3 \, \left( 1 - \frac{\rho_o}{\rho_c} \right) }
				        { \frac{2}{5} \frac{\rho_o}{\rho_c} \alpha - \frac{3}{5} \left( \frac{R_c}{R_p} \right)^5 \left( 1 - \frac{\rho_o}{\rho_c} \right)^2 \frac{\rho_o}{\rho_c} \left( \alpha + \frac{3}{2} \right) \frac{1}{H} }  
	 \:,
\end{split}
\end{equation}
where $H$ acounts for the quantity
\begin{equation}
	H = 1 + \tilde{\mu} - \frac{\rho_o}{\rho_c} 
	    + \frac{3}{2} \frac{\rho_o}{\rho_c} \left( 1 - \frac{\rho_o}{\rho_c} \right) 
	    + \frac{3}{2} \left( \frac{R_c}{R_p} \right)^5 \left( 1 - \frac{\rho_o}{\rho_c} \right)^2 \:.
\end{equation}

Since we have weak constraints on the viscoelastic parameters of giant planet cores (Guillot 2005), we thus have to explore a large range of values.
Fig.~\ref{fig:Graph_Qeff_rheology} shows the tidal dissipation factor $\hat{Q}_\mrm{eff}$ around the reference values presented in Sect.~\ref{sect:params}, expanding the range up to about $\pm$ 2-4 orders of magnitude for $G$ and $\eta$.
In the middle region (inside the blue rectangle on Fig.~\ref{fig:Graph_Qeff_rheology}), where $\eta$ and $G$ correspond to the reference values, the dissipation factor $\hat{Q}_\mrm{eff}$ of Saturn (resp. Jupiter) may reach values of the order of $10^{3}$ (resp. $10^{4}$), and in the whole field it varies up to $10^{20}$.

From Fig.~\ref{fig:Graph_Qeff_rheology}, we deduce that the tidal dissipation of the core may reach the values observed for Jupiter (Lainey et al. 2009) and Saturn (Lainey et al. 2012) assuming that Jupiter's core (resp. Saturn's core) has a radius $34.92$\% (resp. $18.72$\%) larger than this of Guillot 1998 (resp. Hubbard 2009). 

Therefore, we can evaluate the real part of the second order Love numbers ${\tilde{k}_2^F(R_c)}$ and ${\tilde{k}_2^F(R_p)}$, accounting respectively for the deformation of the core's and planet's surface, for parameters whose values are compatible with the tidal dissipation observations (Lainey et al., 2009-2012).
For Jupiter and Saturn, in this order, assuming  that ${R_c = \{ 0.170 , 0.260 \} \times R_p}$, ${G = \{ 4.85 , 4.45 \} \times 10^{10} \, \mrm{Pa}}$ and ${\eta = \{ 1.26 , 1.78 \} \times 10^{14} \, \mrm{Pa}\cdot\mrm{s} }$, we obtain that ${\Re \left[\tilde{k}_2^F(R_c) \right] = \{ 3.21 , 3.31 \} }$ and ${\Re \left[\tilde{k}_2^F(R_p) \right] = \{ 1.37 , 0.24 \} }$.
These estimations at the planet's surface can be compared to the value of Gavrilov \& Zharkov (1977) of $k_2 = 0.379$ for Jupiter and $k_2 = 0.341$ for Saturn obtained for stratified models.
As discussed in the aforementioned paper, the differences between both evaluations are linked to the degree of stratification: the more the planet interior is stratified, the smaller the second order Love number (we recall that the second order Love number of a homogeneous fluid planet is ${3}/{2}$).

\subsection{Dependence of tidal dissipation on the size of the core and the tidal frequency}

For Fig.~\ref{fig:Graph_Q1}, the values of the viscoelastic parameters $G$ and $\eta$, and of the core size $R_c$ result from Fig.~\ref{fig:Graph_Qeff_rheology}: we chose them so as the tidal dissipation factor $\hat{Q}$ reaches the observed values of Lainey et al. (2009-2012) on the condition that the rheological parameters stand in the more realistic domain defined by the lowest and highest value of the ice and rock viscoelasticities taken as reference.
Taking into account the global dissipation values obtained by Lainey et al. 2009 \& 2012 for Jupiter and Saturn, we may deduce some constraints on the viscoelastic parameters and also the size of the core (looking at the red dashed line in Fig.~\ref{fig:Graph_Qeff_rheology}).
We thus will assume that they take values allowing such a dissipation:
\begin{itemize}
	\item we first chose a core slightly larger than that assumed until now: ${R_c = 0.170 \times R_p}$ for Jupiter, and ${R_c = 0.260 \times R_p}$ for Saturn,
	\item we then fixed the value of the shear modulus $G$ to the lowest value needed to reach the observed tidal dissipation of Lainey et al. (2009-2012), {\it i.e.} ${G = 4.85 \times 10^{10} \, \mrm{Pa}}$ for Jupiter, and ${G = 4.45 \times 10^{10} \, \mrm{Pa}}$ for Saturn,
	\item e finally searched the {\it more realistic} value of the viscosity which corresponds to the observed tidal dissipations of Jupiter and Saturn: ${\eta = 1.26 \times 10^{14} \, \mrm{Pa}\cdot\mrm{s}}$ for Jupiter, and ${\eta = 1.78 \times 10^{14} \, \mrm{Pa}\cdot\mrm{s}}$ for Saturn.
\end{itemize}

Fig.~\ref{fig:Graph_Q1} explores, for the present model, the dependence of ${\hat{Q}_\mrm{eff}}$ on the pulsation $\omega$ and the size of the core $R_c$ normalized by the size of the planet $R_p$.
With these parameters, the figure indicates that Saturn dissipates slightly more than ten times more than Jupiter, since ${\left(R_c\right)_\mrm{Saturn}>\left(R_c\right)_\mrm{Jupiter}}$ and ${\left(\rho_c\right)_\mrm{Saturn}< \left(\rho_c\right)_\mrm{Jupiter}}$.
In the range of tidal frequencies of Jupiter's and Saturn's satellites (${2.25 \times 10^{-4} ~\mrm{rad}\cdot\mrm{s}^{-1} ~<~ \omega~<~2.95 \times 10^{-4} ~\mrm{rad}\cdot\mrm{s}^{-1}}$, Lainey et al. 2009-2012), the effective dissipation factor ${\hat{Q}_\mrm{eff}}$ keeps the same order of magnitude.
However, it strongly depends on the size of the core, as it may loose up to $6$ orders of magnitude between a coreless planet and a fully-solid one.
Note that for a given core (so that $M_c$, $R_c$ and then $\rho_c$ are fixed) and a given mass of the planet $M_p$, the density of the fluid envelope $\rho_o$, which varies with its height ${R_p - R_c}$, can not exceed $\rho_c$.
Since:
\begin{equation}
	\rho_o \left( R_c \right) = \frac{M_p - M_c}{4/3 \, \pi \, \left( R_p^3 - R_c^3 \right)} \:,
\end{equation}
this condition gives a limit for the core size:
\begin{equation}
	\left( \frac{R_c}{R_p} \right)_\mrm{sup}= \left( \frac{M_c}{M_p} \right)^{1/3} \:.
\end{equation}

In 2004, Ogilvie \& Lin also studied tidal dissipation in giant planets, and particulary the tidal dissipation resulting from the excitation of inertial waves in the convective region by the tidal potential for rotating giant planets with an elastic solid core.
They obtained a decrease of the quality factor $Q$ of one order of magnitude considering the dynamical tide (due to inertial modes) compared to the equilibrium one: from ${Q = 10^6}$ to $Q=10^5$, but it is not efficient enough to explain the observed tidal dissipations in Jupiter or Saturn which are of 1-2 orders of magnitude higher (Lainey et al. 2009-2012).
Moreover, they showed that the dissipation resulting from the resonance between fluid tide and inertial modes highly depends on the tidal frequency in the range of inertial waves, as do also the coreless models (Wu 2005).
This disagrees with the weak frequency-dependence obtained from astrometry (Lainey et al. 2012).

By discussing the size of the core, Goodman \& Lackner (2009) got a higher value of the quality factor $Q$, in the range $10^7-10^8 \times \left( 0.2 \, R_p / R_c \right)^5$ which disagrees with the observed value of the tidal dissipation of Saturn (see Fig.~\ref{fig:Graph_Qeff_rheology}).

The present two-layer model proposes an alternative process to reach such a high dissipation with a smooth frequency-dependence of $\hat{Q}$.
But the uncertainties on the structural and rheological parameters does not allow us to firmly conclude that the tidal dissipation of the core can explain by its own the tidal dissipation observed in giant planets of our Solar System by Lainey et al. (2009-2012).

Through these expressions of the tidal dissipation closely linked to the internal structure of the planet and its rheological properties, we are now able to derive the equations of the dynamical evolution of the system with explicit dependence on the tidal frequency.

\subsection{Comparison with previous work of Dermott}

The difference beetwen our study and the work of Dermott (1979) lies in the treatment of the tidal dissipation.

Dermott draws his expression from an evolution scenario of Saturnian and Jovian systems.
He assumes that the satellites of Jupiter and Saturn were formed ${4.5 \times 10^9}$ ago, and that their semi-major axis have changed by 10 \% since their formation, with a stable resonance of the main satellites of Jupiter (Io, Europa and Ganymede) since their formation but a young resonance of the satellites Mimas and Thetys of Saturn.
He also assumes an average value of the tidal dissipation, independent of amplitude, frequency and time.
All these assumptions lead Dermott (1979) to a tidal dissipation factor $Q$ that only depends on the mass $M_c$, the size $R_c$, the elasticity $\mu$ of the core and a dimensionless coefficient $K$ characteristic of the evolution scenario of the planet.
In particular, his tidal factor $Q$ is directly proportional to $R_c^5$ (Eq.~27 of Dermott 1979), so that the tidal dissipation gets lower as the core size increases (see Fig.~4 of Dermott 1979), while one should expect an opposite behaviour. 

Instead, our model is constructed on physical considerations of the internal structure and properties of the core.
In particular, we derived our tidal dissipation factor $\hat{Q}$ with no assumption on the evolution of the Jovian and Saturnian systems.
To do so, we used the correspondence principle of Biot (1954).
This allowed us to obtain an expression of the tidal dissipation factor valid for any rheological model of planets' cores.
Moreover, our expression \eqref{Q'} of $\hat{Q}$ not only depends on the mass $M_c$, the size $R_c$ and the elasticity $\mu$ of the core, but also on the tidal frequency $\omega$ and the viscosity $\eta$.
We notice, in particular, that tidal dissipation increases when the size of the core increases, contrary to Dermott's strange result.

\section{Equations of the dynamical evolution}

\begin{figure}[!htb]
\centering
 \includegraphics[width=\linewidth] {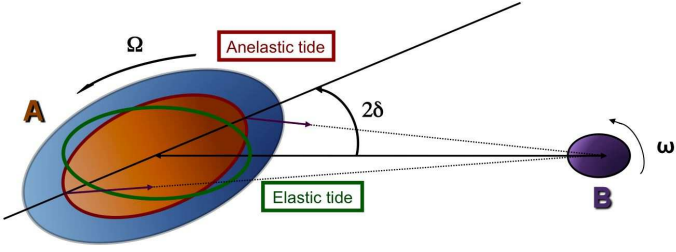}
 \caption{
 Tidal interaction involving a solid body. 
 Body B exerts a tidal force on body A, which adjusts itself with a phase lag $2 \, \delta$, because of internal friction in the anelastic core.
 This adjustment may be split in an adiabatic component, corresponding to the elastic deformation, in phase with the tide, and a dissipative one, resulting from the viscous internal frictions, which is in quadrature.
 }
 \label{fig:Bulge}
\end{figure}

Mass redistribution due to the tide generates a tidal torque of non-zero average which induces an exchange of angular momentum between the orbital motion and the rotation of each component.
As shown in MLP09 \& Remus et al. (2012), this tidal torque is proportional to the tidal dissipation ratio $\displaystyle {{\left|\tilde{k}^F_2\right|}/{Q}}$ (see also Correia \& Laskar 2003, Correia et al. 2003, and Murray \& Dermott 2000).
Note that for a perfectly elastic material, the core will be elongated in the direction of the line of centers, inducing a torque, $\displaystyle {\int_{\mathcal V} \, \mbf r \wedge \left( \rho_c \nab U \right) \, \rm{d} \mathcal{V}}$, with periodic variations of zero average, so that no secular exchanges of angular momentum will be possible (see Zahn 1966a and Remus et al. 2012). 
On the other hand, if the core is anelastic, the deformation of the core resulting from the equilibrium adjustment presents a time delay $\Delta t$ with respect to the tidal forcing, which may be measured also by the tidal lag angle ${2 \delta_l}$ or equivalently by the quality factor $Q$ (see Ferraz-Mello et al. 2008 or Efroimsky \& Williams 2009):
\begin{equation}
	\tan \left[ \Delta t \times \sigma_{2,m,p,q}   \right] =
	\tan \left[ 2 \delta \left( \sigma_{2,m,p,q} \right)  \right] = 
	\frac{1}{Q\left( \sigma_{2,m,p,q} \right)} \:.
\end{equation}
Thus, the tidal bulge is no more aligned with the line of centers, as shown in Fig.~\ref{fig:Bulge}.

The resulting tidal angle is at the origin of a torque of non-zero average which causes exchange of spin and orbital angular momentum between the components of the system.

The evolution of the semi-major axis $a$, of the eccentricity $e$, of the inclination $I$, of the obliquity $\varepsilon$ and of the angular velocity $\Omega$ ($\bar{I}_A$ denotes the moment of inertia of $A$), is governed by the following equations, established in MLP09 and Remus et al. (2012):

\begin{multline}
	\frac{\mrm d \left( \bar{I}_A \, \Omega \right) }{\mrm dt}
		= - \frac{8\pi}{5} \frac{\mcal G M_{B}^{2} R_\mrm{eq}^5}{a^6} \\
			\times \sum_{(m,j,p,q) \in \mathbb{I}}
			\left\{ \frac{\left|\tilde{k}^F_2(R_p,\sigma_{2,m,p,q})\right|}{\hat{Q}_\mrm{eff}(\sigma_{2,m,p,q})} \left[ \mcal H_{2,m,j,p,q}(e,I,\varepsilon) \right]^2 \right\} \:,
\label{Evol_Omega}
\end{multline}
\vspace*{-0.4cm}
\begin{multline}
	\bar{I}_A \, \Omega \, \frac{\mrm d \left( \cos\varepsilon \right)}{\mrm dt}
		=  \frac{4\pi}{5} \frac{\mcal G M_{B}^{2} R_\mrm{eq}^5}{a^6} \\
			\times \sum_{(m,j,p,q) \in \mathbb{I}}
			\left\{ (j+2\cos\varepsilon) \, \frac{\left|\tilde{k}^F_2(R_p,\sigma_{2,m,p,q})\right|}{\hat{Q}_\mrm{eff}(\sigma_{2,m,p,q})} \left[ \mcal H_{2,m,j,p,q}(e,I,\varepsilon) \right]^2 \right\} \:,
\label{Evol_epsilon}
\end{multline}
\vspace*{-0.4cm}
\begin{multline}
	\frac{1}{a} \, \frac{\mrm da}{\mrm dt} 
		= -\frac{2}{n}\frac{4\pi}{5}\frac{\mcal G M_{B} R_\mrm{eq}^5}{a^8} \\
			\times \sum_{(m,j,p,q) \in \mathbb{I}}
			\left\{(2-2p+q) \, \frac{\left|\tilde{k}^F_2(R_p,\sigma_{2,m,p,q})\right|}{\hat{Q}_\mrm{eff}(\sigma_{2,m,p,q})} \left[ \mcal H_{2,m,j,p,q}(e,I,\varepsilon) \right]^2
			\right\} \:,
\label{Evol_a}
\end{multline}
\vspace*{-0.4cm}
\begin{multline}
	\frac{1}{e} \, \frac{{\rm d}e}{{\rm d}t} 
		= - \frac{1}{n}\frac{1-e^2}{e^2}\frac{4\pi}{5}\frac{\mcal G M_{B} R_\mrm{eq}^5}{a^8} \\ 
	 	\times \sum_{(m,j,p,q) \in \mathbb{I}}
	 		\left\{ \vphantom{\frac{\left|\tilde{k}^F_2\right|}{\hat{Q}_e(\sigma_q)}} \left[ (2-2p) \, \left( 1 - \frac{1}{\sqrt{1-e^2}} \right) + q \right] \right. \\ 
	 			\times \left. \frac{\left|\tilde{k}^F_2(R_p,\sigma_{2,m,p,q})\right|}{\hat{Q}_\mrm{eff}(\sigma_{2,m,p,q})} \left[\mcal H_{2,m,j,p,q}(e,I,\varepsilon) \right] ^2 \vphantom{\left( \frac{1}{\sqrt{e^2}} \right)} \right\} \:,
\label{Evol_e}
\end{multline}
\vspace*{-0.4cm}
\begin{multline}
	\frac{\mrm d \left( \cos I \right)}{\mrm dt}
		=  \frac{1}{n} \, \frac{1}{\sqrt{1-e^2}} \, \frac{4\pi}{5} \frac{\mcal G M_{B}^{2} R_\mrm{eq}^5}{a^8} \\
			\times \sum_{(m,j,p,q) \in \mathbb{I}}
			\left\{ \vphantom{\frac{\left|\tilde{k}^F_2\right|}{\hat{Q}}} \left[ j+(2q-2)\cos I \right] \right. \\ 
			\times \left. \frac{\left|\tilde{k}^F_2(R_p,\sigma_{2,m,p,q})\right|}{\hat{Q}_\mrm{eff}(\sigma_{2,m,p,q})} \left[ \mcal H_{2,m,j,p,q}(e,I,\varepsilon) \right]^2 \right\} \:,
\label{Evol_i}
\end{multline}

where the functions $\mcal H_{m,j,p,q}(e,I,\varepsilon)$ are expressed in terms of $d_{j,m}^{2}(\varepsilon)$, $F_{2,j,p}(I)$, and $G_{2,p,q}(e)$, which are defined in Sect.~\ref{section:potU}
\begin{multline}
	\mcal H_{2,m,j,p,q}(e,I,\varepsilon) = \sqrt{ \frac{5}{4\pi} \frac{(2-|j|)!}{(2+|j|)!} } \\
		\times d_{j,m}^{2}(\varepsilon) \, F_{2,j,p}(I) \, G_{2,p,q}(e) \:,
\end{multline}
and $R_\mrm{eq}$ designates the equatorial radius of body $A$.

From these equations one may derive the characteristic times of synchronization, circularization and spin alignment:
\begin{flalign}
	&\frac{1}{t_{\rm sync}}
		= - \frac{1}{\Omega-n}\frac{\mrm d \Omega}{\mrm dt} 
		= - \frac{1}{\bar{I}_A \, \left(\Omega-n\right)}\frac{\mrm d \left( \bar{I}_A \, \Omega \right) }{\mrm dt} \:, \\
	&\frac{1}{t_{\rm circ}}
		= - \frac{1}{e} \, \frac{{\rm d}e}{{\rm d}t} \:, \\
	&\frac{1}{t_{\rm align_A}}
		= - \frac{1}{\varepsilon} \, \frac{{\rm d}\varepsilon}{{\rm d}t} 
		= \frac{1}{\varepsilon \, \sin\varepsilon} \, \frac{{\rm d} \left(\cos\varepsilon\right)}{{\rm d}t} \:, \\
	&\frac{1}{t_{\rm align_{Orb}}}
		= - \frac{1}{I} \, \frac{{\rm d}I}{{\rm d}t}
		= \frac{1}{I \, \sin I} \, \frac{{\rm d} \left(\cos I \right)}{{\rm d}t} \:.
\end{flalign}

\section{Conclusion}

To conclude, the purpose of this work was to study the tidal dissipation in a two-layer planet consisting in a rocky/icy core and a fluid envelope, as one expects to be the case in Jupiter, Saturn, and many extrasolar planets. 
We considered the most general configuration where the perturber (star or satellite) is moving on an elliptical and inclined orbit around the planet which rotates on an inclined axis. 
We expanded the tidal displacement in Fourier series and spherical harmonics, each term of the expansion having a radial part proportional to that of the corresponding term of the tidal potential, which depends on the eccentricity, the inclination and the obliquity.
We followed the method by Dermott (1979) to derive the modified Love numbers $h _2^F$ and $k _2^F$ accounting for the tidal deformation at the boundary of the solid core. 
Like Dermott, we made the simplifying assumption that core and envelope have a constant density. 
Then, generalizing the results of his work invoking the correspondence principle, we obtained the tidal dissipation rate of the core expressed by $| k _2^F|/\hat Q$, $\hat Q$ being the quality factor. 
That ratio depends on the tidal frequency and on the rheological properties of the core; unlike Dermott we made no assumption on the formation history of the system.
As mentioned in Sect.~\ref{sect:params} the rheological properties of planetary cores are still quite uncertain. 
However, taking plausible values for the viscoelastic parameters $G$ and $\eta$, we obtain a tidal dissipation which may be much higher than for a fully fluid planet and weakly frequency-dependent.
Under these assumptions, we find that the low value of ${Q = (1.682 \pm 0.540) \times 10^3}$, determined by Lainey et al. (2012) and needed by Charnoz et al. (2011) to explain the formation of all mid-size moons of Saturn from the rings, can be reached taking into account the tidal dissipation of Saturn's core.
In the same way, the dissipation in Jupiter's core may explain the value of the $Q$-factor determined by Lainey et al. (2009), i.e. ${Q = (3.56 \pm 0.56) \times 10^4}$.
But to do so, we need to assume a core in Jupiter and Saturn slightly larger than those of Guillot (1998) and Hubbard et al. (2009).
But we recall that in our model  the density was assumed to be piecewise constant. In the future, we shall consider a non constant density profile, to evaluate the impact on our results of a realistic density stratification.
Moreover, there are much uncertainties on the determination of the core size in giant planets as Jupiter and Saturn, so that we need more constraints on the system formation by core accretion (Pollack et al. 1996) and differenciation resulting from the internal structure evolution (Nettelmann 2011).
Furthermore, seismology seems to offer an interesting way to improve our knowledge of giant planets interiors (Gaulme et al. 2011).

To conclude, the purpose of this paper was to study tidal dissipation in the solid parts of a simple model of two-layer planet; we show how this mechanism may be powerful.
The results derived here are general in the sense that no specific rheological model has been assumed. 
However, considering the lack of constraints on the rheology of giant planets cores, we have chosen the simplest Maxwell model to illustrate the tidal dissipation.  

This work represents thus a first step for further numerical investigations in more realistic cases.

\section*{\normalsize Acknowledgments}
The authors are grateful to the referee for his/her remarks and suggestions.
They also thank G. Tobie for fruitful discussions during this work and T. Guillot for providing numerical models of Jupiter and Saturn interiors. This work was supported in part by the Programme National de Plan\'etologie (CNRS/INSU), the EMERGENCE-UPMC project EME0911, and the CNRS programme {\it Physique th\'eorique et ses interfaces}.



\renewcommand{\appendixpagename}{}
\begin{appendices}
\section*{Appendix: elastic system}

In Sect.~\ref{subsection:dyn_eq} we gave the system of equations \eqref{syst_elastic}, with its boundary conditions \eqref{CL_elastic}, governing an elastic planet under tidal perturbation.
Considering that such perturbations are of small order of magnitude compared to the hydrostatic equilibrium, we proposed in Sect.~\ref{subsection:dyn_eq_lin} a method to linearize the system \eqref{syst_elastic_lin}.
Thus, assuming the expansion \eqref{syst:spheric_exp} of all quantities in spherical harmonics, we obtain the following system of equations governing the scalar radial parts of these expansions (Alterman et al. 1959; Takeuchi \& Saito 1972):
\begin{subequations}
\label{syst:rad_funct}
\begin{flalign}
	\dot{y}^m_1 =& - \frac{2 \left(K-\frac{2}{3} \mu \right)}{K+\frac{4}{3}\mu} \, \frac{y^m_1}{r} + \frac{1}{K+\frac{4}{3}\mu} \, y^m_2 
					+ \frac{6 \left(K-\frac{2}{3} \mu \right)}{K+\frac{4}{3}\mu} \, \frac{y^m_3}{r} \:, \\
	\dot{y}^m_2 =& \left[ -4 \, \rho_0 \, g_s \, \frac{r}{R} + \frac{12 \mu K}{\left( K+\frac{4}{3}\mu \right) r} \right]  \, \frac{y^m_1}{r} 
					- \frac{4 \mu K}{\left( K+\frac{4}{3}\mu \right) r}  \, \frac{y^m_2}{r} \nonumber\\
				 & + 6 \left[ \rho_0 \, g_s \, \frac{r}{R} - \frac{6 \mu K}{\left( K+\frac{4}{3}\mu \right) r} \right]  \, \frac{y^m_3}{r}
					+6 \, \frac{y^m_4}{r}
					- \rho_0 y^m_6 \:, \\
	\dot{y}^m_3 =& - \frac{y^m_1}{r} + \frac{y^m_3}{r} + \frac{y^m_4}{\mu} \:, \\
	\dot{y}_4^m =& \left[ \rho_0 \, g_s \, \frac{r}{R} - \frac{6 \mu K}{\left( K+\frac{4}{3}\mu \right) r} \right]  \, \frac{y^m_1}{r} \nonumber\\
				 & + \frac{2 \mu \left[ 11 \, \left(K-\frac{2}{3} \mu \right) + 10 \, \mu \right]}
							{\left( K+\frac{4}{3}\mu \right) r}  \, \frac{y^m_3}{r} \:, \\
	\dot{y}^m_5 =& 4 \pi \mcal G \rho_0 y^m_1 +y^m_6 \:, \\
	\dot{y}^m_6 =& -24 \pi \mcal G \rho_0 \frac{y^m_3}{r} + \frac{6}{r} \frac{y^m_5}{r} - 2 \, \frac{y^m_6}{r} \:.
\end{flalign}
\end{subequations}

Solutions of \eqref{syst:rad_funct} are given by \eqref{syst:solutions}.

\end{appendices}


\begin{thebibliography}{99}

\bibitem[Alterman et al.(1959)]{1959RSPSA.252...80A} Alterman, Z., Jarosch, H., \& Pekeris, C.~L.\ 1959, Royal Society of London Proceedings Series A, 252, 80
\bibitem[Biot(1954)]{1954JAP....25.1385B} Biot, M.~A.\ 1954, Journal of Applied Physics, 25, 1385
\bibitem[Charnoz et al.(2011)]{2011Icar..216..535C} Charnoz, S., Crida, A., Castillo-Rogez, J.~C., et al.\ 2011, \icarus, 216, 535 
\bibitem[Charette \& Smith (2010)]{2010Ocea...104..106C} Charette, M.A., \& Smith, W.H.F.\ 2010, Oceanography, 104, 106
\bibitem[Chree(1896)]{1896CaPhTr...16..14C} Chree, C. \ 1896, Cambridge Phil. Trans., 16, 14
\bibitem[Correia et al.(2003)]{2003Icar..163....1C} Correia, A.~C.~M., Laskar, J., \& de Surgy, O.~N.\ 2003, Icarus, 163, 1
\bibitem[Correia \& Laskar(2003)]{2003Icar..163...24C} Correia, A.~C.~M., \& Laskar, J.\ 2003, Icarus, 163, 24 
\bibitem[Dahlen et al.(1999)]{1999PhT....52h..61D} Dahlen, F.~A., Tromp, J., \& Lay, T.\ 1999, Physics Today, 52, 61
\bibitem[Dermott(1979)]{1979Icar...37..310D} Dermott, S.~F.\ 1979, \icarus, 37, 310
\bibitem[]{} Efroimsky, M., \& Williams, J. G. 2009, Celestial Mechanics and Dynamical Astronomy, 104, 257
\bibitem[]{} Ferraz-Mello, S., Rodriguez, A., \& Hussmann, H.\ 2008, Celestial Mechanics and Dynamical Astronomy, 101, 171
\bibitem[Gaulme et al.(2011)]{2011A&A...531A.104G} Gaulme, P., Schmider, F.-X., Gay, J., Guillot, T., \& Jacob, C.\ 2011, \aap, 531, A104
\bibitem[Gavrilov \& Zharkov(1977)]{1977Icar...32..443G} Gavrilov, S.~V., \& Zharkov, V.~N.\ 1977, \icarus, 32, 443  
\bibitem[Goodman \& Lackner(2009)]{2009ApJ...696.2054G} Goodman, J., \& Lackner, C.\ 2009, \apj, 696, 2054
\bibitem[Greff-Lefftz et al.(2005)]{2005CeMDA..93..113G} Greff-Lefftz, M., M{\'e}tivier, L., \& Legros, H.\ 2005, Celestial Mechanics and Dynamical Astronomy, 93, 113 
\bibitem[Guillot et al.(1995)]{1995ApJ...450..463G} Guillot, T., Chabrier, G., Gautier, D., \& Morel, P.\ 1995, \apj, 450, 463
\bibitem[Guillot \& Morel(1995)]{1995A&AS..109..109G} Guillot, T., \& Morel, P.\ 1995, \aaps, 109, 109
\bibitem[Guillot(1999)]{1999P&SS...47.1183G} Guillot, T.\ 1999, \planss, 47, 1183
\bibitem[Guillot(2005)]{2005AREPS..33..493G} Guillot, T.\ 2005, Annual Review of Earth and Planetary Sciences, 33, 493 
\bibitem[Henning et al.(2009)]{2009ApJ...707.1000H} Henning, W.~G., O'Connell, R.~J., \& Sasselov, D.~D.\ 2009, \apj, 707, 1000
\bibitem[Hubbard et al.(2009)]{2009sfch.book...75H} Hubbard, W.~B., Dougherty, M.~K., Gautier, D., \& Jacobson, R.\ 2009, Saturn from Cassini-Huygens, 75
\bibitem[Karato(2011)]{2011Icar..212...14K} Karato, S.-I.\ 2011, \icarus, 212, 14 
\bibitem[Kaula(1962)]{1962AJ.....67..300K} Kaula, W.~M.\ 1962, \aj, 67, 300
\bibitem[Lainey et al.(2009)]{2009Natur.459..957L} Lainey, V., Arlot, J.-E., Karatekin, {\"O}., \& van Hoolst, T.\ 2009, \nat, 459, 957
\bibitem[Lainey et al.(2012)]{Lainey...ApJ..2012} Lainey, V., Karatekin, {\"O}., Desmars, J., Charnoz, S., Arlot, J.-E., Emelyanov, N., Le Poncin-Lafitte, C., Mathis, S., Remus, F., Tobie, G., \& Zahn, J.-P. 2012, submited to \apj
\bibitem[Lambeck(1980)]{1980esvr.book.....L} Lambeck, K.\ 1980, The Earth's Variable Rotation (Cambridge University Press)
\bibitem[Love(1911)]{1911spge.book.....L} Love, A.~E.~H.\ 1911, Some Problems of Geodynamics (Cambridge University Press)
\bibitem[Mathis \& Le Poncin-Lafitte(2009)]{2009A&A...497..889M} Mathis, S., \& Le Poncin-Lafitte, C.\ 2009, \aap, 497, 889
\bibitem[Mathis \& Zahn(2005)]{2005A&A...440..653M} Mathis, S., \& Zahn, J.-P.\ 2005, \aap, 440, 653
\bibitem[Melchior(1966)]{1966tet.book.....L} Melchior, P.\ 1966, The Earth Tides (Pergamon, New York)
\bibitem[Militzer et al.(2008)]{2008ApJ...688L..45M} Militzer, B., Hubbard, W.~B., Vorberger, J., Tamblyn, I., \& Bonev, S.~A.\ 2008, \apjl, 688, L45
\bibitem[Militzer \& Hubbard(2009)]{2009Ap&SS.322..129M} Militzer, B., \& Hubbard, W.~B.\ 2009, \apss, 322, 129 
\bibitem[Murray \& Dermott(2000)]{2000ssd..book.....M} Murray, C.~D., \& Dermott, S.~F.\ 2000, Solar System Dynamics (Cambridge University Press)
\bibitem[Nettelmann et al.(2008)]{2008ApJ...683.1217N} Nettelmann, N., Holst, B., Kietzmann, A., et al.\ 2008, \apj, 683, 1217 
\bibitem[Nettelmann(2011)]{2011Ap&SS.336...47N} Nettelmann, N.\ 2011, \apss, 336, 47
\bibitem[Ogilvie \& Lin(2004)]{2004ApJ...610..477O} Ogilvie, G.~I., \& Lin, D.~N.~C.\ 2004, \apj, 610, 477
\bibitem[Ogilvie \& Lin(2007)]{2007ApJ...661.1180O} Ogilvie, G.~I., \& Lin, D.~N.~C.\ 2007, \apj, 661, 1180
\bibitem[Ogilvie(2009)]{2009MNRAS.396..794O} Ogilvie, G.~I.\ 2009, \mnras, 396, 794 
\bibitem[Peale \& Cassen (1978)]{} Peale, S.J., \& Cassen, P.\ 1978, \icarus, 36, 245
\bibitem[Pollack et al.(1996)]{1996Icar..124...62P} Pollack, J.~B., Hubickyj, O., Bodenheimer, P., et al.\ 1996, \icarus, 124, 62
\bibitem[Remus et al.(2012)]{Remus...A&A..2012} Remus, F., Mathis, S. \& Zahn, J.-P. 2012, submited to \aap
\bibitem[Rieutord(1987)]{1987GApFD..39..163R} Rieutord, M.\ 1987, Geophysical and Astrophysical Fluid Dynamics, 39, 163
\bibitem[Ross \& Schubert(1986)]{1986LPSC...16..447R} Ross, M., \& Schubert, G.\ 1986, Lunar and Planetary Science Conference Proceedings, 16, 447 
\bibitem[Santos \& et al.(2007)]{2007jena.confE...5S} Santos, N.~C., \& et al.\ 2007, JENAM-2007, ''Our Non-Stable Universe''
\bibitem[Saumon \& Guillot(2004)]{2004ApJ...609.1170S} Saumon, D., \& Guillot, T.\ 2004, \apj, 609, 1170
\bibitem[Takeuchi \& Saito(1972)]{} Takeuchi, H., and M. Saito 1972, Seismic surface waves, Methods Comput. Phys., 11, 217–295
\bibitem[Thomson(1863)]{} Thomson, W. (Lord Kelvin) 1863, Dynamical problems regarding elastic spheroidal shells, and On the rigidity of the Earth, Phil. Trans. Roy. Soc. London 153, 573-616.
\bibitem[Tobie(2003)]{2003PhD} Tobie, G.\ 2003, Impact du chauffage de mar\'ee sur l\textquoteright\'evolution g\'eodynamique d\textquoteright Europe et de Titan, PhD thesis, Universit\'e Paris 7 - Denis Diderot 
\bibitem[Tobie et al.(2005)]{2005Icar..177..534T} Tobie, G., Mocquet, A., \& Sotin, C.\ 2005, \icarus, 177, 534 
\bibitem[Wu(2005)]{2005ApJ...635..688W} Wu, Y.\ 2005, \apj, 635, 688
\bibitem[Yoder(1995)]{1995Icar..117..250Y} Yoder, C.~F.\ 1995, \icarus, 117, 250 
\bibitem[Zahn(1966a)]{1966AnAp...29..313Z} Zahn, J.-P.\ 1966a, Annales d'Astrophysique, 29, 313
\bibitem[Zahn(1966b)]{1966AnAp...29..489Z} Zahn, J.~P.\ 1966b, Annales d'Astrophysique, 29, 489
\bibitem[Zahn(1977)]{1977A&A....57..383Z} Zahn, J.-P.\ 1977, \aap, 57, 383 

\end{thebibliography}
\end{document}